\documentclass[twocolumn]{aastex631}

\usepackage{comment}
\usepackage{amsmath}	\usepackage{amssymb}	\usepackage{upgreek}
\usepackage{tikz}
\usetikzlibrary{matrix}
\usepackage{graphicx}
\usepackage{bm}
\usepackage{multirow}
\usepackage{orcidlink}
\usepackage{hyperref}
\usepackage{cleveref}

\newcommand{\ha}{H$\alpha$}

\newcommand{\hb}{H$\beta$}
\newcommand{\mgii}{Mg{\sc ii}}
\newcommand{\civ}{C{\sc iv}}

\newcommand{\kmps}{$\rm km\,s^{-1}$}

\newcommand{  \niii     }{\ifmmode \textrm{N}\,\textsc{iii} \else N\,\textsc{iii}\fi}
\newcommand{  \NIII     }{\ifmmode \textrm{N}\,\textsc{iii}\,\lambda4640 \else N\,\textsc{iii}\,$\lambda4640$\fi}
\newcommand{  \OIIIbf   }{\ifmmode \textrm{O}\,\textsc{iii}\,\lambda3133 \else O\,\textsc{iii}\,$\lambda3133$\fi}

\begin{document}

\title{The Large Sky Area Multi-object Fiber Spectroscopic Telescope (LAMOST) Quasar Survey: Quasar Properties from Data Release 10 to 12}

\correspondingauthor{}
\email{lyubing@pku.edu.cn, wuxb@pku.edu.cn}

\author[0000-0001-8879-368X]{Bing Lyu}
\affiliation{Kavli Institute for Astronomy and Astrophysics, Peking University, Beijing 100871, People's Republic of China }

\author[0000-0002-7350-6913]{Xue-Bing Wu}
\affiliation{Kavli Institute for Astronomy and Astrophysics, Peking University, Beijing 100871, People's Republic of China }
\affiliation{Department of Astronomy, School of Physics, Peking University, Beijing, 100871, People's Republic of China \\}

\author[0000-0002-8402-3722]{Jun-Jie Jin}
\affiliation{Key Laboratory of Optical Astronomy, National Astronomical Observatories, Chinese Academy of Sciences, Beijing 100012, People's Republic of China}

\author[0000-0002-0759-0504]{Yuming Fu}
\affil{Leiden Observatory, Leiden University, Einsteinweg 55, 2333 CC Leiden, The Netherlands}
\affil{Kapteyn Astronomical Institute, University of Groningen, P.O. Box 800, 9700 AV Groningen, The Netherlands}

\author[0009-0005-3823-9302]{Yuxuan Pang}
\affiliation{Kavli Institute for Astronomy and Astrophysics, Peking University, Beijing 100871, People's Republic of China }
\affiliation{Department of Astronomy, School of Physics, Peking University, Beijing, 100871, People's Republic of China \\}

\author[0000-0001-8803-0738]{Huimei Wang}\affiliation{Kavli Institute for Astronomy and Astrophysics, Peking University, Beijing 100871, People's Republic of China }
\affiliation{Department of Astronomy, School of Physics, Peking University, Beijing, 100871, People's Republic of China \\}

\author[0000-0002-0792-2353]{Rui Zhu}\affiliation{Kavli Institute for Astronomy and Astrophysics, Peking University, Beijing 100871, People's Republic of China }
\affiliation{Department of Astronomy, School of Physics, Peking University, Beijing, 100871, People's Republic of China \\}

\author[0000-0002-9728-1552]{Su Yao}
\affiliation{National Astronomical Observatories, Chinese Academy of Sciences, Beijing 100012, People’s Republic of China}
\author[0000-0001-9312-4640]{Yan-Li Ai}
\affiliation{College of Engineering Physics, Shenzhen Technology University, Shenzhen 518118, China}
\affiliation{Shenzhen Key Laboratory of Ultraintense Laser and Advanced Material Technology, Shenzhen 518118, China}
\author[0000-0002-6610-5265]{Yan-xia Zhang}
\affiliation{National Astronomical Observatories, Chinese Academy of Sciences, Beijing 100012, People’s Republic of China}
\author{Hai-long Yuan}
\affiliation{National Astronomical Observatories, Chinese Academy of Sciences, Beijing 100012, People’s Republic of China}
\author[0009-0003-3066-2830]{Zhi-ying Huo}
\affiliation{National Astronomical Observatories, Chinese Academy of Sciences, Beijing 100012, People’s Republic of China}

\begin{abstract}
We present the quasar catalog from Data Releases 10 to 12 of the Large Sky Area Multi-Object Fiber Spectroscopic Telescope (LAMOST) Quasar Survey, comprising quasars observed between September 2021 and June 2024. We robustly identified $11,346$ quasars, of which $5,386$ are newly discovered objects not present in the Million Quasars catalog. This release brings the total number of quasars identified by the 12-year LAMOST survey to $67,521$, of which $29,513$ are newly discovered. While the absolute flux calibration for LAMOST quasar spectra from Data Releases 6 to 9 was previously performed using the SDSS/PanSTARRS1 multi-band photometric data, the inherent variability of quasars can affect the flux accuracy. To address this limitation, we recalibrated the LAMOST spectra using (quasi-)simultaneous photometric data from Zwicky Transient Facility (ZTF), which has conducted high-cadence sky monitoring since March 2018. Based on the recalibrated single-epoch spectra, we estimated the emission line fluxes, continuum fluxes, and virial black hole masses. These improved spectra facilitate direct comparison with the spectra of common quasars from the Sloan Digital Sky Survey (SDSS), enabling searches for rare quasars, such as changing-look quasars exhibiting the appearance or disappearance of broad emission lines and broad absorption line quasars. The combined dataset of photometry and multi-epoch spectra will enhance the detections of AGN-related transients, such as Bowen fluorescence flares and extreme variability quasars, thereby improving our understanding of quasar variability.
\end{abstract}


\section{Introduction} \label{sec:intro}

Quasars are the most luminous class of active galactic nuclei (AGNs) in the universe, powered by the accretion of material onto the supermassive black holes (SMBHs) at the center of their host galaxies. Quasars emit multi-wavelength electromagnetic radiation from low-frequency radio to high-energy $\gamma$-ray band \citep{1993ARA&A..31..473A}. Owing to their exceptional luminosities, quasars are detectable at early cosmic epochs, providing a unique approach to trace the early growth phases of SMBHs across cosmic time \citep[e.g.,][]{2023ARA&A..61..373F} and to map the large-scale structure of the early universe \citep[e.g.,][]{2001AJ....122.2850B}. Furthermore, quasars serve as critical laboratories for investigating the co-evolution of SMBHs and their host galaxies \citep[e.g.,][]{2013ARA&A..51..511K}, and as backlights to probe the distribution of interstellar and intergalactic medium \citep[ISM and IGM;][]{2007ApJ...655..735H}.

Since the discovery of the first quasar in 1963 \citep{1963Natur.197.1040S}, millions of quasars have been found under great efforts during the past decades \citep[e.g.,][]{2023OJAp....6E..49F}. Quasars are distinguished from normal galaxies and stars by some typical features, which include their high luminosities outshining entire galaxies, spectra featuring broad emission lines, characteristic spectral energy distribution (SED), and detectable variability in brightness over timescales ranging from hours to years \citep[e.g.,][]{2016AJ....151...24A,2022ApJ...941..191L,2023MNRAS.518.1531L}.

A widely adopted strategy for selecting quasar candidates leverages their distinct multi-wavelength photometric properties. Quasars at $z <$ 2.2 exhibit strong UV excess that can be distinguished from stars in the color-color and color-magnitude diagrams as demonstrated by large-scale surveys like the Sloan Digital Sky Survey (SDSS; \citealt{2011ApJS..194...45S,2012A&A...548A..66P,2018A&A...613A..51P}) and the Two-Degree Fields (2dF) Quasars Redshift Survey \citep{2000MNRAS.317.1014B}. However, the completeness and efficiency of such optical color selection methods degrade significantly at $2.2 < z < 3.5$, especially for quasars at $z = 2.7$, as the stellar objects occupy a similar region in the color-color diagram \citep{2007AJ....134..102S} and are difficult to distinguish. The inclusion of near-infrared (NIR) photometry of K-band data from the UK Infrared Telescope (UKIRT) Infrared Deep Sky Survey \citep[UKIDSS;][]{2007MNRAS.379.1599L} provides an efficient way to discover quasars at $2.2 < z< 3.5$ due to the characteristic NIR excess for quasars relative to stars \citep[][]{2000MNRAS.312..827W}. Furthermore, \citet{2010MNRAS.406.1583W} and \citet{2012AJ....144...49W} propose effective quasar selection criteria with the combination of optical photometry and infrared data from the \textit{Wide-field Infrared Survey Explorer} \citep[\textit{WISE};][]{2010AJ....140.1868W}, which significantly improve the purity and completeness of quasar catalogs.

Machine learning algorithms have become increasingly instrumental in the efficient selection of quasar candidates from large-scale photometric surveys \citep[e.g.,][]{2012ApJS..199....3R,2019MNRAS.485.4539J,2024MNRAS.527.4677Z}.  Various supervised learning techniques, including random forests, support vector machines, and boosting algorithms, have demonstrated remarkable success in distinguishing quasars from stellar contaminants. A transfer learning XGBoost algorithm is used to select quasars behind the Galactic plane \citep[GPQ,][]{2021ApJS..254....6F,2022ApJS..261...32F}. Besides, an improved Gaia DR3 quasar candidate catalog (CatNorth) is similarly built with data from Gaia, Pan-STARRS1, and CatWISE2020 \citep[][]{2024ApJS..271...54F}, which combines the multi-color photometric data and proper motion information. These photometrically selected quasar candidates (e.g., from GPQ and CatNorth catalogs) serve as the primary input for the spectroscopic follow-up observations in the Large Sky Area Multi-Object Fiber Spectroscopic Telescope \citep[LAMOST,][]{1996ApOpt..35.5155W,2004ChJAA...4....1S} phase III quasar survey. While the photometric selection is highly efficient for candidate identifications, the definitive confirmation and precise redshift measurement of quasars ultimately depend on spectroscopic observations. With its large aperture and multi-fiber capability, LAMOST is well-suited for this confirmation task.

While the LAMOST survey provides a vast spectroscopic dataset, a recognized limitation is the inherent challenge of achieving the precise absolute flux calibration for its spectra. This imprecision can hinder the detailed spectral analysis, such as accurate measurements of the emission line fluxes, continuum slopes, and their temporal variations, thereby limiting the scope of subsequent scientific investigations. To overcome this critical issue, we have implemented a recalibration of the LAMOST spectra using high-cadence photometric data from the Zwicky Transient Facility \citep[ZTF; e.g.,][]{2019PASP..131a8002B} . This calibration enhancement enables more accurate measurements of the spectral features and continuum properties. The combination of precisely calibrated spectra and multi-epoch photometric light curves from ZTF provides a powerful opportunity to systematically mine the LAMOST quasar sample for rare and extreme phenomena. Our specific science goals focus on the identification and characterization of Bowen Fluorescence Flares \citep[BFFs; e.g.,][]{2023ApJ...953...32M}, extreme variability quasars \citep[EVQs; e.g.,][]{2018ApJ...854..160R,2024ApJ...963....7R}, broad absorption line quasars \citep[BALs; e.g.,][]{2024MNRAS.532.3669F}, and changing-look quasars \citep[CLQs; e.g.,][]{2015ApJ...800..144L,2018ApJ...862..109Y,2025ApJ...986..160D}, etc.

This paper presents the quasar catalog from the LAMOST quasar survey from Data Release 10 (DR10) to Data Release 12 (DR12), with observations conducted between September 2021 and June 2024. It is the fifth part in a series of LAMOST quasar survey publications, after Data Release 1 \citep[DR1,][hereafter Paper I]{2016AJ....151...24A}, Data Release 2 and 3 \citep[DR2 and DR3,][hereafter Paper II]{2018AJ....155..189D},  Data Release 4 and 5 \citep[DR4 and DR5,][hereafter Paper III]{2019ApJS..240....6Y} and Data Release from 6 to 9 \citep[DR6 to DR9,][hereafter Paper IV]{2023ApJS..265...25J}. 

The outline of the paper is as follows. In \autoref{sec:outline}, we provide an overview of the candidate selection strategy, the spectroscopic survey, and the quasar identification methodology. \autoref{sec:spectra} details the absolute flux calibration of the spectra, the procedures for spectral fitting, and the estimation of virial black hole masses ($M_{\rm BH}$). The description of the quasar catalog and key parameters is presented in \autoref{sec:catalog}. A brief discussion and summary of the results are presented in \autoref{sec:dis} and \autoref{sec:sum}. Throughout this work, we adopt a flat $\Lambda-$CDM cosmology with $H_0$ = 70 km s$^{-1}$ Mpc $^{-1}$, $\Omega_{m}$ = 0.27, and $\Omega_{\Lambda} = 0.73$.

\section{Survey Outline} \label{sec:outline}
LAMOST, also known as Guoshoujing Telescope, is a quasi-meridian reflecting Schmidt telescope located at Xinglong Observatory in China \citep{1996ApOpt..35.5155W,2004ChJAA...4....1S,2012RAA....12.1197C}. It has an effective aperture of 3.6-4.9 meters and a 5-degree field of view. The telescope is equipped with 4,000 optical fibers (each with a diameter of 3.3 arcsec) mounted on the focal plane, which are connected to 16 spectrographs. For the quasar survey, observations were carried out in the low-resolution mode (R $\sim$ 1000-2000). Each spectrum covers a blue channel (3700-5900 $\rm \AA$) and a red channel (5700-9000 $\rm \AA$), with a 200 $\rm \AA$ overlap. Individual exposures were typically set to 90 minutes, often split into three sub-exposures, adjusted according to the observational conditions and target brightness. 

Following its commissioning phase (2009–2010) and a pilot survey in 2011 \citep{2012RAA....12.1243L}, LAMOST began its regular survey in September 2012. The quasar survey was conducted under the LAMOST Experiment for Galactic Understanding and Exploration (LEGUE) and the LAMOST ExtraGAlactic Survey (LEGAS) project \citep{2012RAA....12..723Z}. From DR1 to DR9, the LAMOST quasar survey has identified more than 50,000 quasars by 2021 (see Paper \hyperlink{cite.2023ApJS..265...25J}{IV}).

\subsection{Target Selection} \label{sec:target}
The selection of quasar candidates for the LAMOST quasar survey have been described in \cite{2010MNRAS.406.1583W}, \cite{2012AJ....144...49W}, \cite{2012MNRAS.425.2599P}, and Papers \hyperlink{cite.2016AJ....151...24A}{I}, \hyperlink{cite.2018AJ....155..189D}{II}, \hyperlink{cite.2019ApJS..240....6Y}{III}, and \hyperlink{cite.2023ApJS..265...25J}{IV}. The primary selection of quasar candidates is based on the combination of optical photometric data from SDSS \citep{2012ApJS..203...21A} and mid-infrared data from WISE/UKIDSS. Only point sources are selected to exclude galaxies. The targets should be brighter than $i=$ 20 to ensure a sufficient signal-to-noise (S/N) ratio for the spectra and fainter than $i=$ 16 to avoid fiber saturation and contamination. Most of the quasar candidates are further selected and separated from stars based on the optical-infrared color criteria \citep[SDSS-UKIDSS/WISE; e.g.,][]{2010MNRAS.406.1583W,2012AJ....144...49W}. 

This primary sample is supplemented via cross-matching SDSS photometry data with multi-waveband surveys (e.g., X-ray surveys like XMM-Newton, Chandra, and ROSAT and radio surveys like FIRST and NVSS) and data-mining algorithms such as SVM classifiers \citep{2012MNRAS.425.2599P}, the extreme deconvolution method (XDQSO; \citealt{2011ApJ...729..141B}), and KDE \citep{2009ApJS..180...67R}. More recently, the GPQ candidates catalog and the CatNorth quasar candidate catalog, which are constructed using a transfer learning algorithm, have also been incorporated as additional input sources for the survey \citep[see ][]{2021ApJS..254....6F,2024ApJS..271...54F}.

\subsection{Pipeline for Data Reduction} \label{sec:pipe}
The reduced spectra from the LAMOST survey are accessible through the LAMOST Data Archive Server \footnote{\url{http://www.lamost.org/lmusers/}} (DAS). The raw CCD images undergo initial processing with the LAMOST two-dimensional (2D) pipeline, which performs dark and bias subtraction, flat-fielding, cosmic-ray removal, spectral extraction, sky subtraction, wavelength calibration, merging subexposure, relative flux calibration, and combining blue and red spectra  \citep{2012RAA....12..453S,2015RAA....15.1095L}.  In the final step of the 2D pipeline, A- and F-type stars with high-quality spectra are used as the pseudo-standard stars to calibrate both the blue and red spectrograph arms. The Lick spectral index grid from \citet{1972PASP...84..161R,1998ApJS..116....1T} is adopted to calculate the effective temperatures of these stars. The spectral response curve (SRC) is obtained by dividing the observed continuum by the physical pseudo-continuum for the star and applied to calibrate all other fiber spectra.

\subsection{Quasar Identification} \label{sec:indent}
The one-dimensional (1D) spectra extracted by the 2D pipeline are subsequently processed by the 1D pipeline \citep{2015RAA....15.1095L} and classified into four categories: ``QSO", ``GALAXY", ``STAR", and ``Unknown" \citep[e.g.,][]{2023ApJS..265...25J}. Spectra classified as ``Unknown" are typically the result of poor observational conditions or unstable efficiency of individual fibers. With continuous improvements in the quasar candidate selection methods and the spectral classification pipeline, the fraction of observed candidates ultimately classified as ``QSO" has been increasing across the survey Data Releases ($\rm \sim 14\%$ in Paper \hyperlink{cite.2016AJ....151...24A}{I}, $55.9\%$ in Paper \hyperlink{cite.2018AJ....155..189D}{II}, $62.3\%$ in Paper \hyperlink{cite.2019ApJS..240....6Y}{III}, $77.0\%$ in Paper \hyperlink{cite.2023ApJS..265...25J}{IV}, and {\color{black} $80.0\%$ } in this work). 

The $13,217$ spectra automatically labeled as ``QSO'' by the pipeline were subjected to visual inspection to confirm their identities with the help of a Java-based program ASERA \citep{2013A&C.....3...65Y}.  When the spectrum features match the quasar template, the corresponding object was classified as a quasar and the redshift was determined when at least one typical quasar emission line (e.g., H$\alpha$, H$\beta$, O{\,\sc iii}$\lambda5007$, Mg{\sc ii}, C{\sc iii} and C{\sc iv}) was confidently identified. In cases where only one clear emission line was present and matched, the spectrum was flagged as ``ZWARNING = 1" to indicate a lower confidence in the redshift measurement. We excluded the background quasars within M31/M33 and Galactic-anti-center extension region (GACext) in our final quasar catalog, which will be studied and published elsewhere \citep[e.g.,][]{2010RAA....10..612H,2013AJ....145..159H,2015RAA....15.1438H}. Besides, we visually checked 1946 spectra of quasar candidates from the input catalog of LAMOST that were not classified as ``QSO" by the pipeline. From this supplemental check, we confirmed an additional 417 quasars that the pipeline had missed.

After the multi-step verification process, we visually confirmed a total of {\color{black} $11,346$ } reliable quasars from DR10 to DR12.  Cross-matching with the Million Quasars catalog (Milliquas\footnote{\url{http://www.quasars.org/milliquas.htm}} v8; \citealt{2023OJAp....6E..49F}), which includes quasars from the literature to 30 June 2023 and quasars from the DESI-EDR and SDSS-DR18Q, {\color{black} $5,960$ } quasars were previously known. Consequently, the remaining {\color{black} $5,386$ } are new discoveries by LAMOST. It is noteworthy that $3,227$ quasars in LAMOST DR11 are also found in the contemporaneous SDSS DR18 quasar catalog \citep{2023ApJS..267...44A}. These quasars are considered independent confirmations by LAMOST. The availability of multi-epoch spectroscopic observations for quasars common to both LAMOST and SDSS provides a valuable opportunity to investigate quasar spectral variability on different timescales and to search for rare phenomena such as changing-look AGNs (CLAGNs; e.g., \citealt{2015ApJ...800..144L,2016ApJ...821...33R,2019ApJ...874....8M,2018ApJ...862..109Y,2019ApJ...887...15W,2019ApJ...883L..44G,2024ApJS..270...26G,2025ApJ...986..160D,2025ApJ...980...91Y}), which exhibit dramatic changes in their broad emission lines.  

\begin{figure*}[!htb]
\centering
\includegraphics[page=1,scale=0.6]{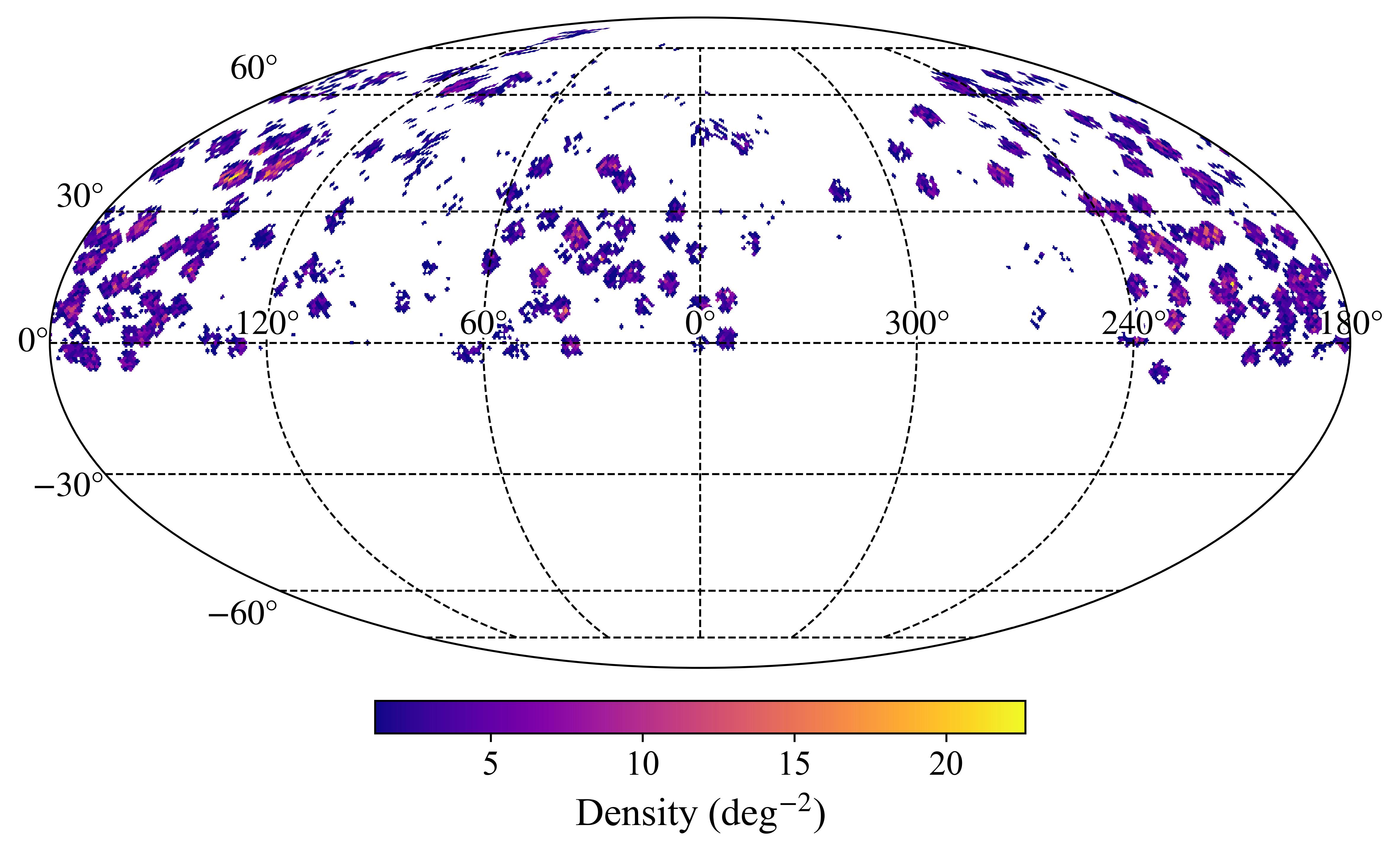}  
            \includegraphics[page=1,scale=0.6]{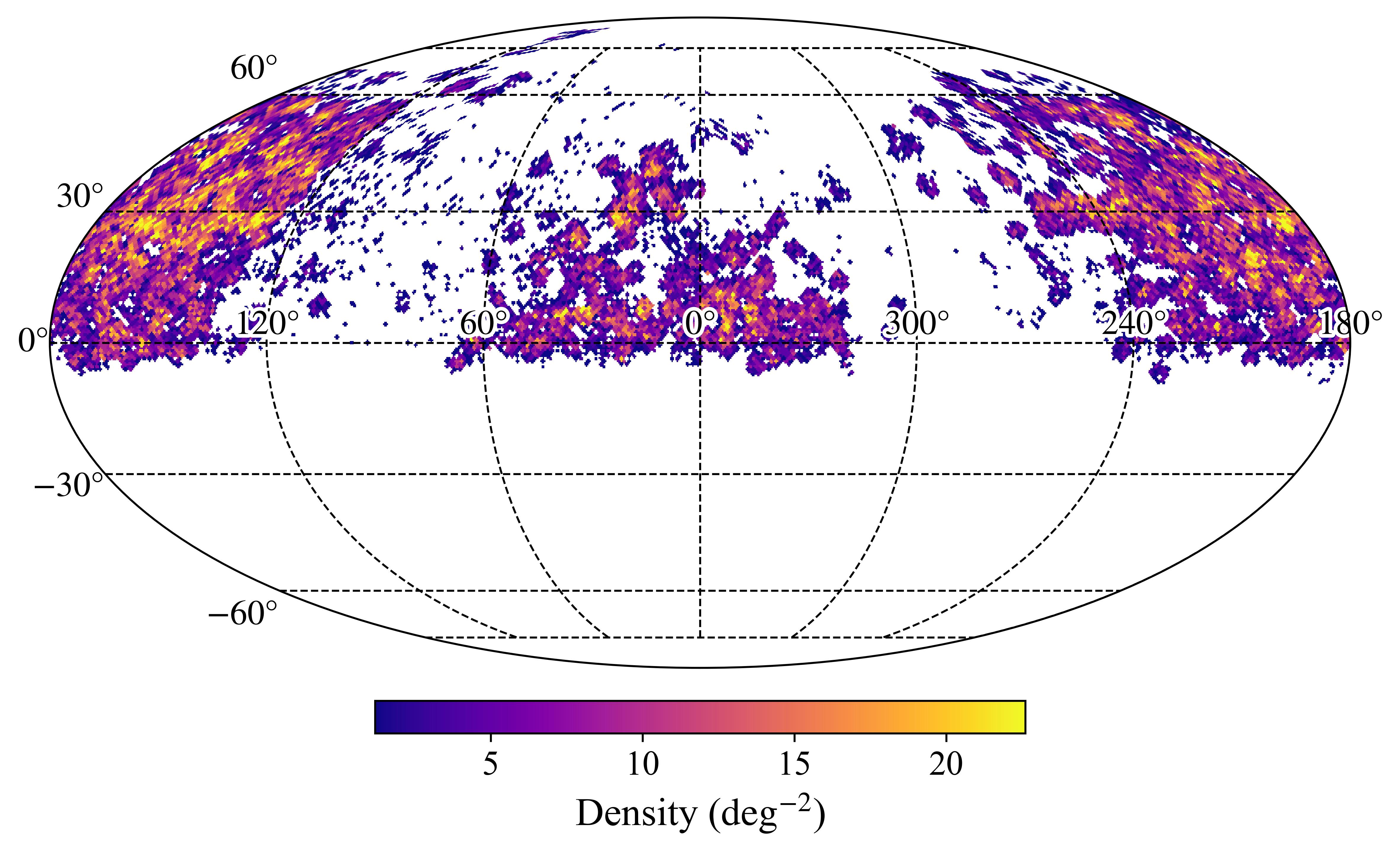}   
		\caption{The HEALPix sky distributions of the quasars identified in LAMOST DR10-12 (upper panel) and DR1-12 (lower panel) are shown in equatorial coordinates with the parameters $\rm N_{side=64}$ and area of 0.839 $\rm deg^{2}$ per pixel.} 
\label{fig:skymap} 
\end{figure*}

\begin{figure}[!htb]
\centering
\includegraphics[page=1,width=0.4\textwidth]{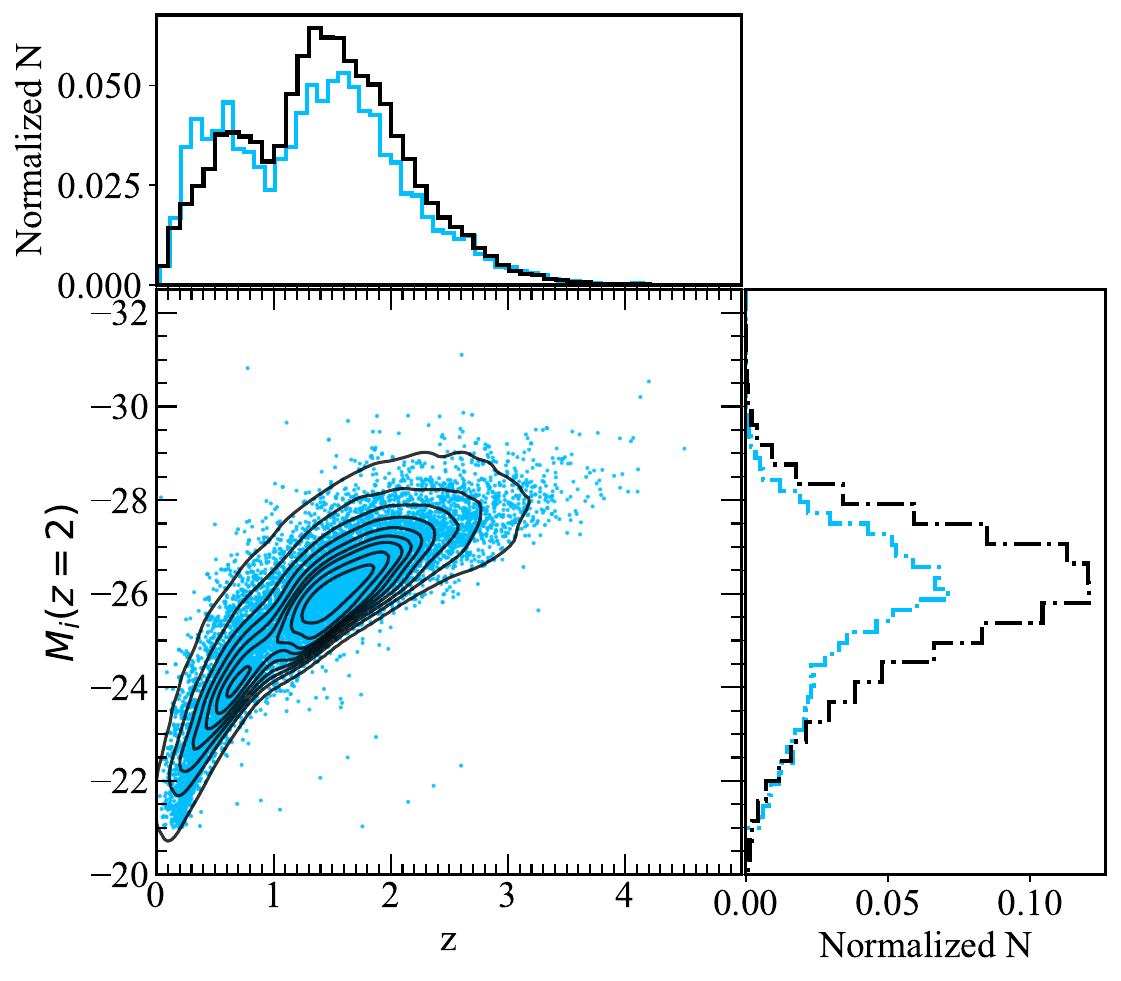}   
\caption{The distribution in the magnitude-redshift space for the visually confirmed quasars for previous (DR1-9) LAMOST quasar survey (black contours) and in DR10-12 (blue). The absolute magnitudes $\rm M_{i} (z=2)$ are normalized at z=2, following the K-correction of \cite{2006AJ....131.2766R}. The upper and right panels show the absolute magnitude and redshift distributions, respectively.}
\label{fig:z_mi} 
\end{figure}

The sky distribution of quasars identified by the LAMOST survey is shown in \autoref{fig:skymap}. A summary of the LAMOST quasar survey results is presented in Table \ref{table:catalog3}. The distribution of the redshift and the K-corrected i-band absolute magnitude ${ \rm M}_{i} (z=2) $, normalized at $z$ = 2 \citep{2006AJ....131.2766R}, is shown in \autoref{fig:z_mi}. A noticeable drop in the number of identifications appears at $ z \sim 1$, consistent with the previous Data Releases (Papers \hyperlink{cite.2016AJ....151...24A}{I},  \hyperlink{cite.2018AJ....155..189D}{II},  \hyperlink{cite.2019ApJS..240....6Y}{III}, and  \hyperlink{cite.2023ApJS..265...25J}{IV}). The inefficient  identification around $ z \sim 1$ occurs since the Mg{\sc ii} emission line moves into the overlapping region of the blue and red channels of the spectrograph.

\begin{figure}[!htb]
\centering
\includegraphics[page=1,width=0.4\textwidth]{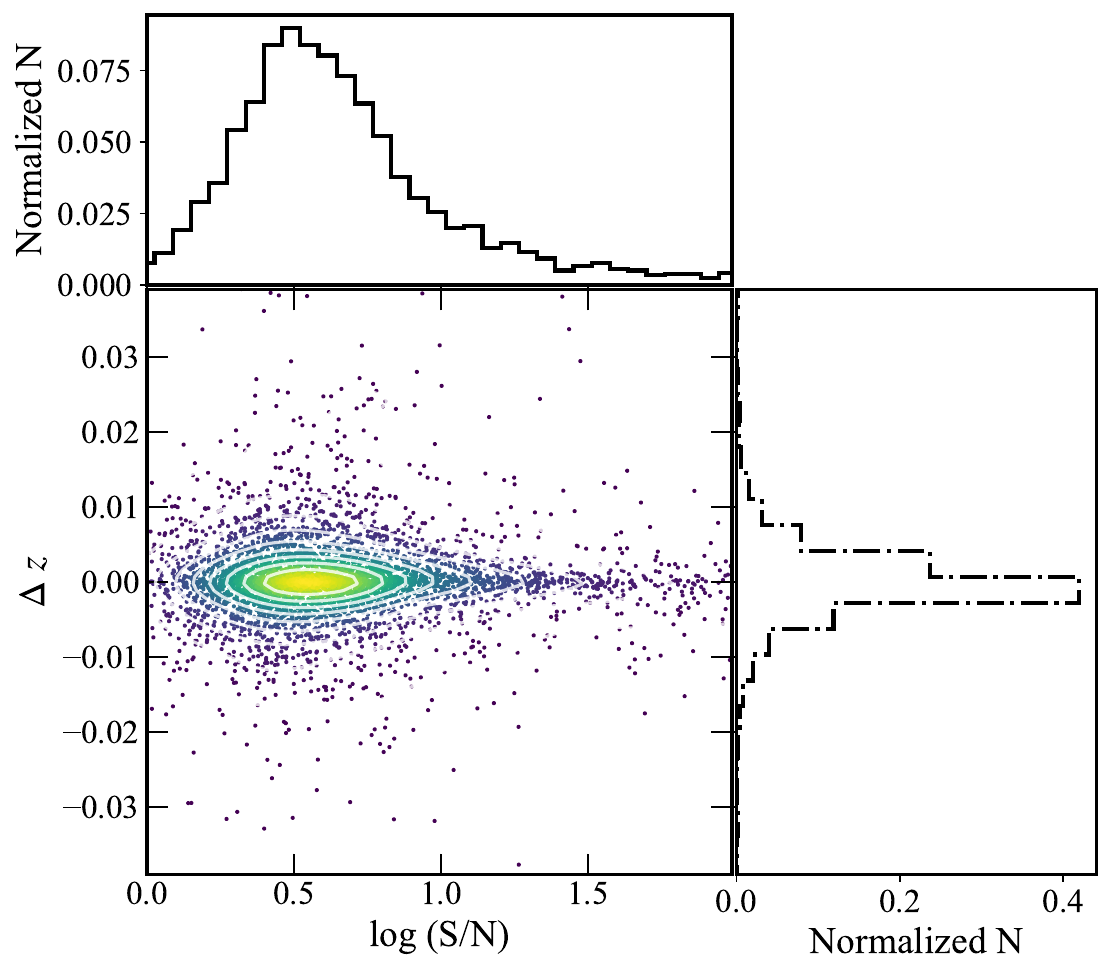}   
		\caption{The distribution of redshift difference ($\Delta z$) for common quasars between this work and SDSS versus LAMOST spectral S/N.}
\label{fig:snr_detlaZ} 
\end{figure}

Among the {\color{black} $5,705$ } sources observed by both SDSS and LAMOST, {\color{black} $199$} of them have a redshift difference ($ \Delta z = z_{\rm LAMOST} - z_{\rm SDSS}$) greater than 0.1. The dominant cause of this discrepancy is the misidentification of emission lines in LAMOST spectra with low S/N. As shown in \autoref{fig:snr_detlaZ}, the redshift difference increases systematically as the S/N decreases. Another factor is that we estimated the redshift based on the strongest typical emission line, while the SDSS values are derived from more sophisticated approaches such as principal component analysis (PCA) or using the Mg{\sc ii} emission line \citep{2018A&A...613A..51P}.

\begin{figure*}[!htb]
\centering
\begin{tikzpicture}
    \matrix[matrix of nodes]{
    \includegraphics[width=0.33\textwidth]{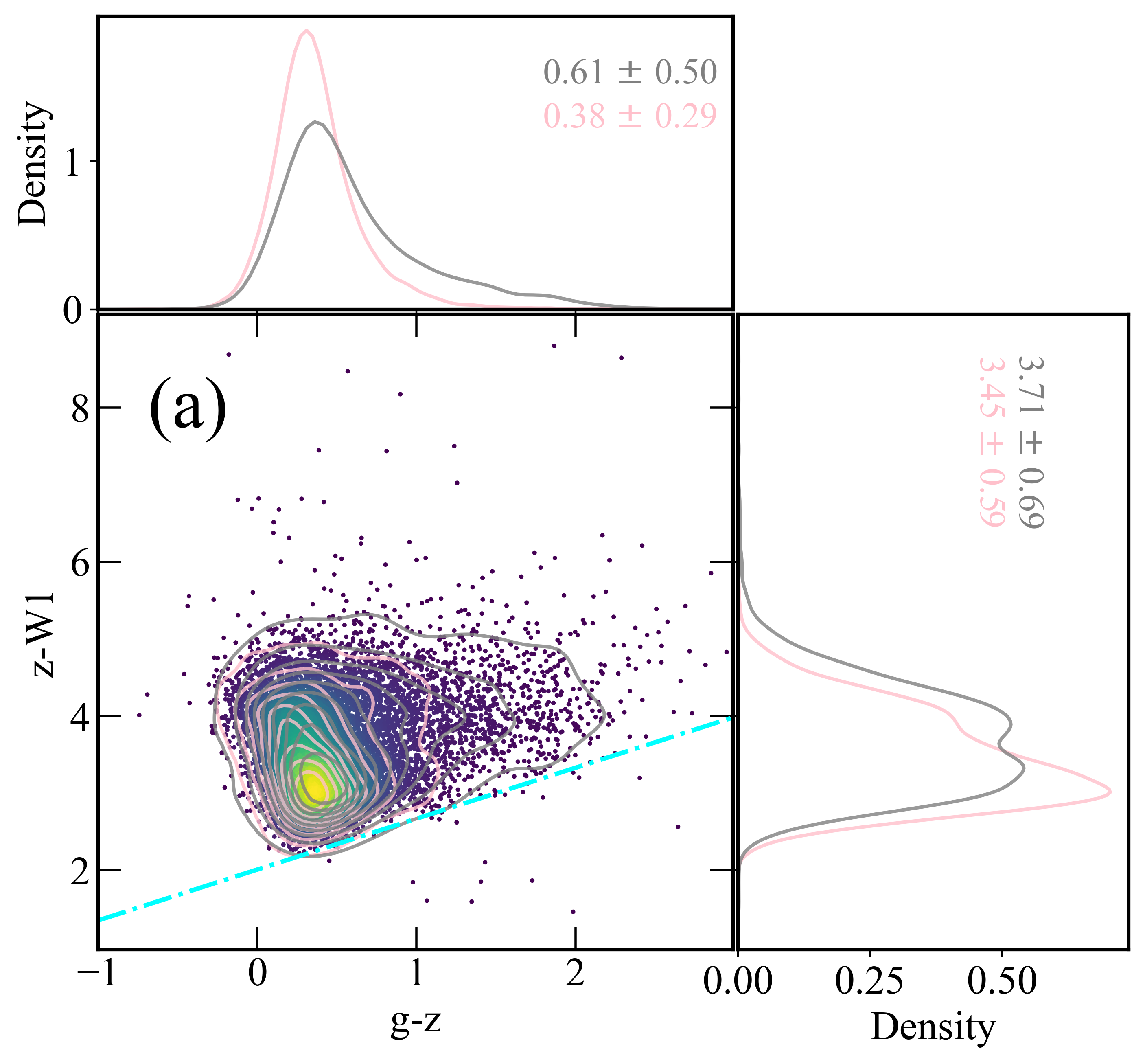} &
    \includegraphics[width=0.33\textwidth]{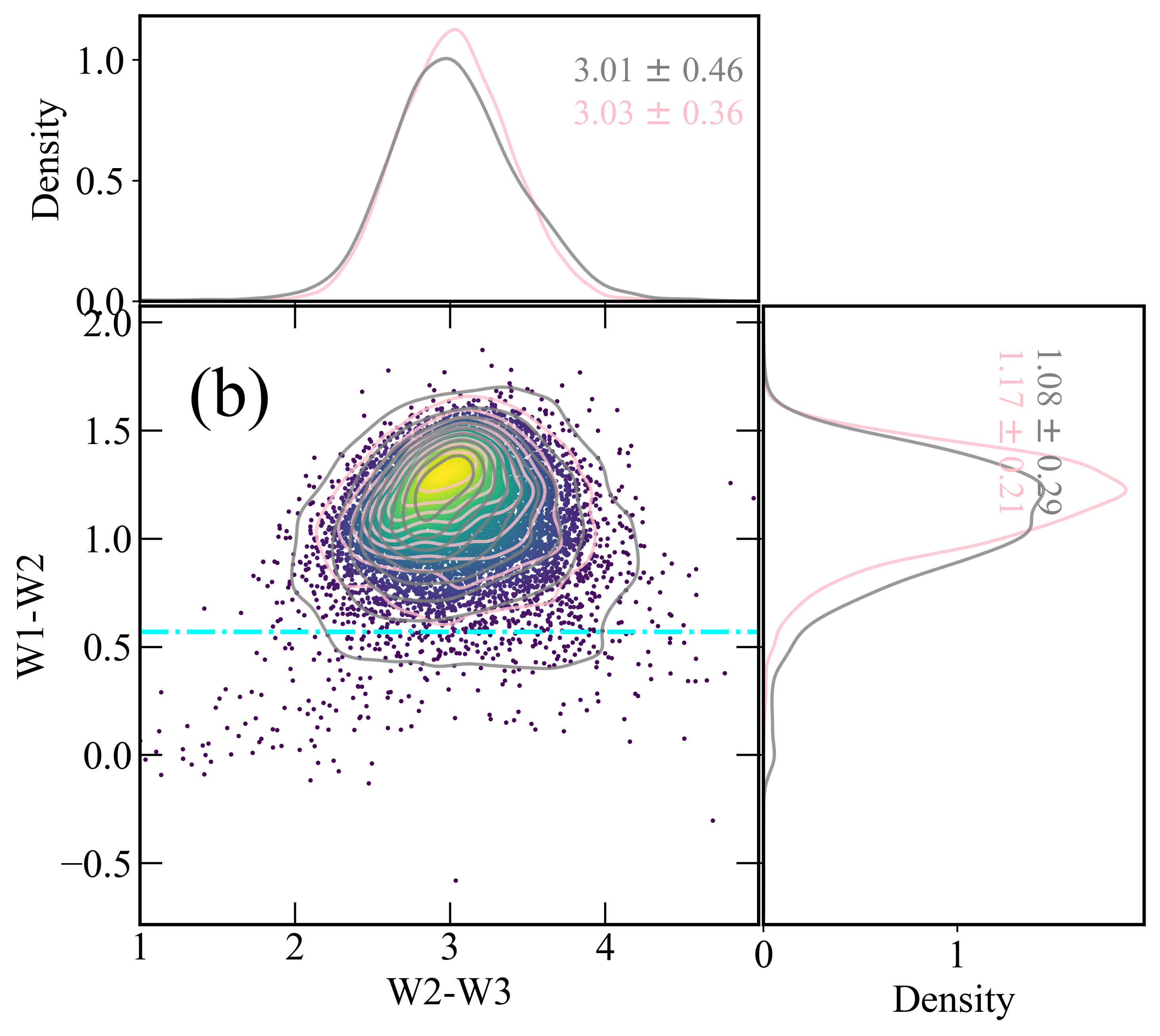} \\ 
     \includegraphics[width=0.33\textwidth]{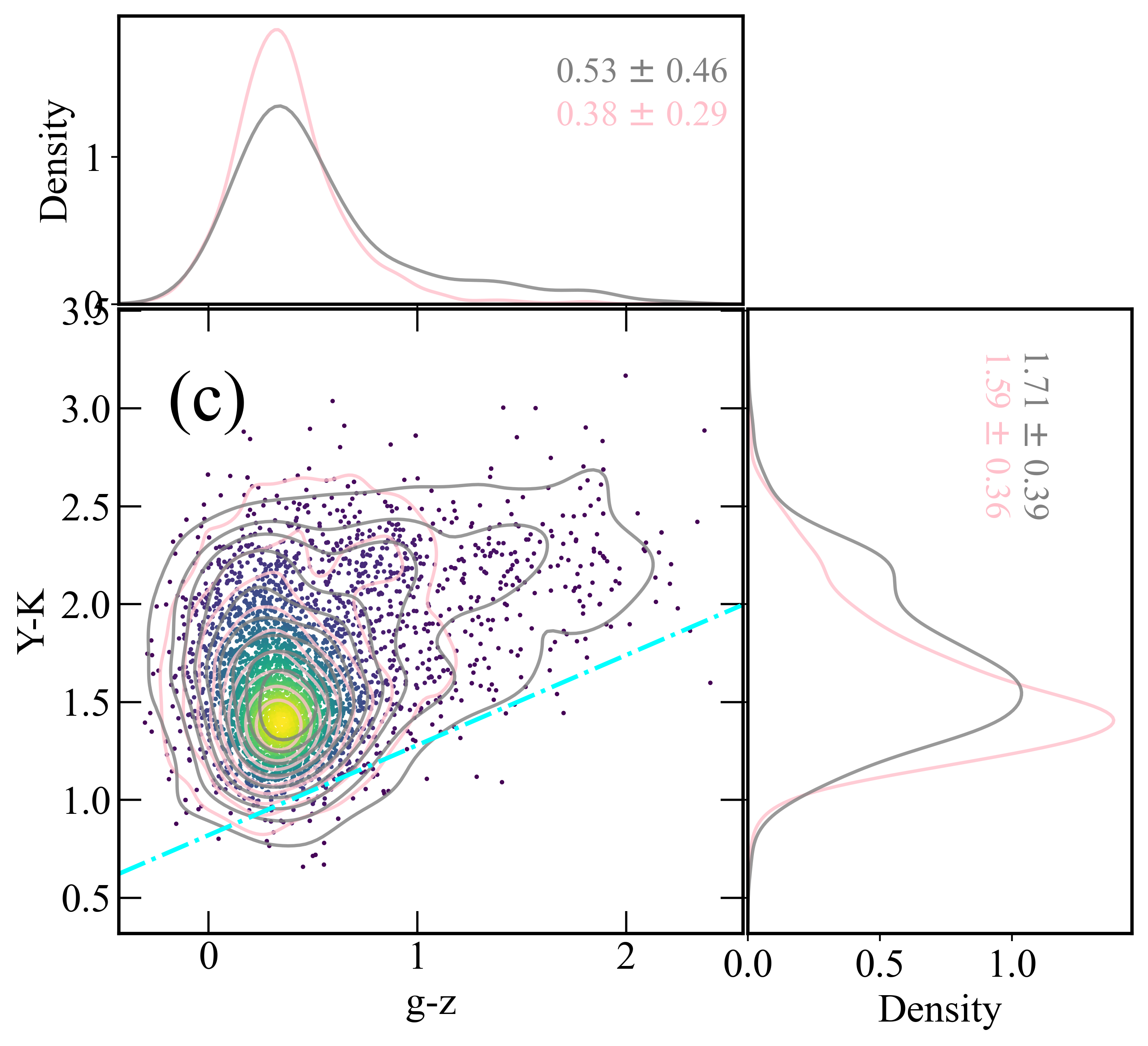} &
 \includegraphics[width=0.33\textwidth]{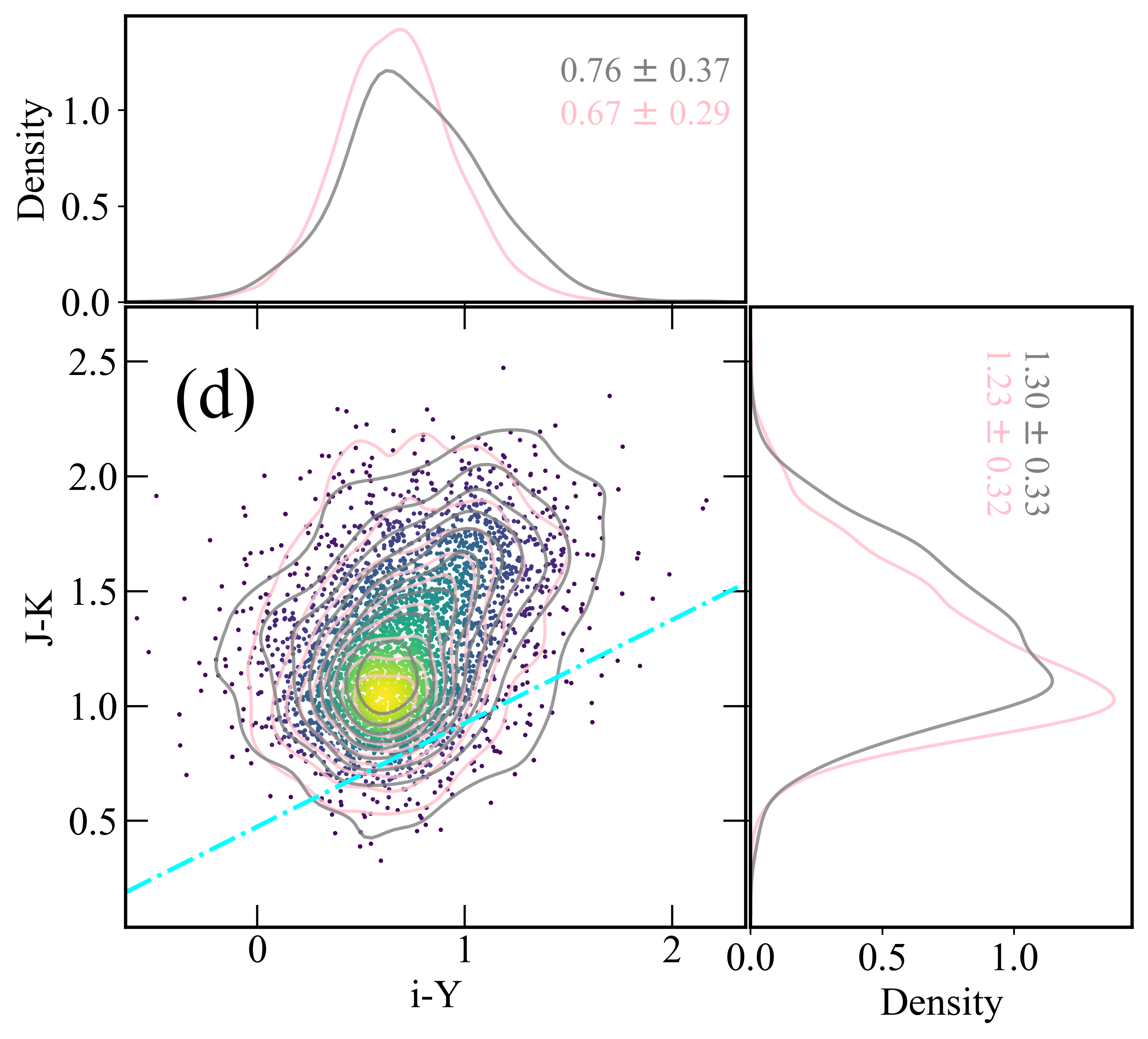} &\\ 
    };
\end{tikzpicture}
\caption{The distributions of LAMOST quasars in the SDSS-WISE/UKIDSS color diagram. The WISE and UKIDSS magnitudes are in Vega magnitudes. The SDSS magnitudes in the panels (a) and (b) are plotted in AB magnitudes. The dash-dotted lines indicate the criteria previously used in the LAMOST QSO survey \citep{2010MNRAS.406.1583W,2012AJ....144...49W}.The contours in pink show the distribution for common quasars between this work and Milliquas, while the contours in gray show the distribution for unique quasars identified in this work. The mean ($\mu$) and dispersion ($\sigma$) of each distribution are tabulated in corresponding plots. }
\label{fig:color}
\end{figure*}

The SDSS-WISE/UKIDSS color-color distributions of the identified quasars are shown in \autoref{fig:color}.  More than 99\% of the confirmed sources lie within the expected selection regions defined by optical-infrared color criteria (see upper-left panel in \autoref{fig:color}), demonstrating the effectiveness of the color-based selection.  Quasars uniquely identified by LAMOST are slightly redder than those common quasars to both LAMOST and SDSS, which is expected since the LAMSOT candidate selection incorporates the additional infrared band data, improving the recovery of redder quasars that may be missed by purely optical selection. 

\begin{table}[]
	\centering
	\caption{The summary of the results of the LAMOST quasars survey up to now.}
	\scriptsize
\begin{tabular}{lll lllc}
		\hline 
		\hline 
		&Papers \hyperlink{cite.2016AJ....151...24A}{I}& \hyperlink{cite.2018AJ....155..189D}{II} & \hyperlink{cite.2019ApJS..240....6Y}{III} &  \hyperlink{cite.2023ApJS..265...25J}{IV} &This Work&Total\\
		\hline 
		Total&3,921&19,935&19,253&13,066&     $11,346$   &    67,521          \\Known&2,741&11,835&11,091&6,381& $5,960$ & 38,008                 \\Independent&1,180&12,126&11,458& 7,102& 8,427 &  40,293       \\New&1,180&8,100&8,162&6,685 & $5,386$ &   29,513          \\\hline 
	\end{tabular}
	\label{table:catalog3}
\end{table}

\section{SPECTRAL ANALYSIS} \label{sec:spectra} 
We briefly outline our spectral analysis procedure, which includes the absolute flux calibration using the ZTF photometry, spectroscopic fitting, measurements of typical quasar emission lines, and the corresponding virial BH mass ($M_{\rm{BH}})$.

\subsection{Absolute Flux Calibration} \label{sec:abscali}
\begin{figure}[!htb]
\centering
\includegraphics[page=1,width=0.4\textwidth]{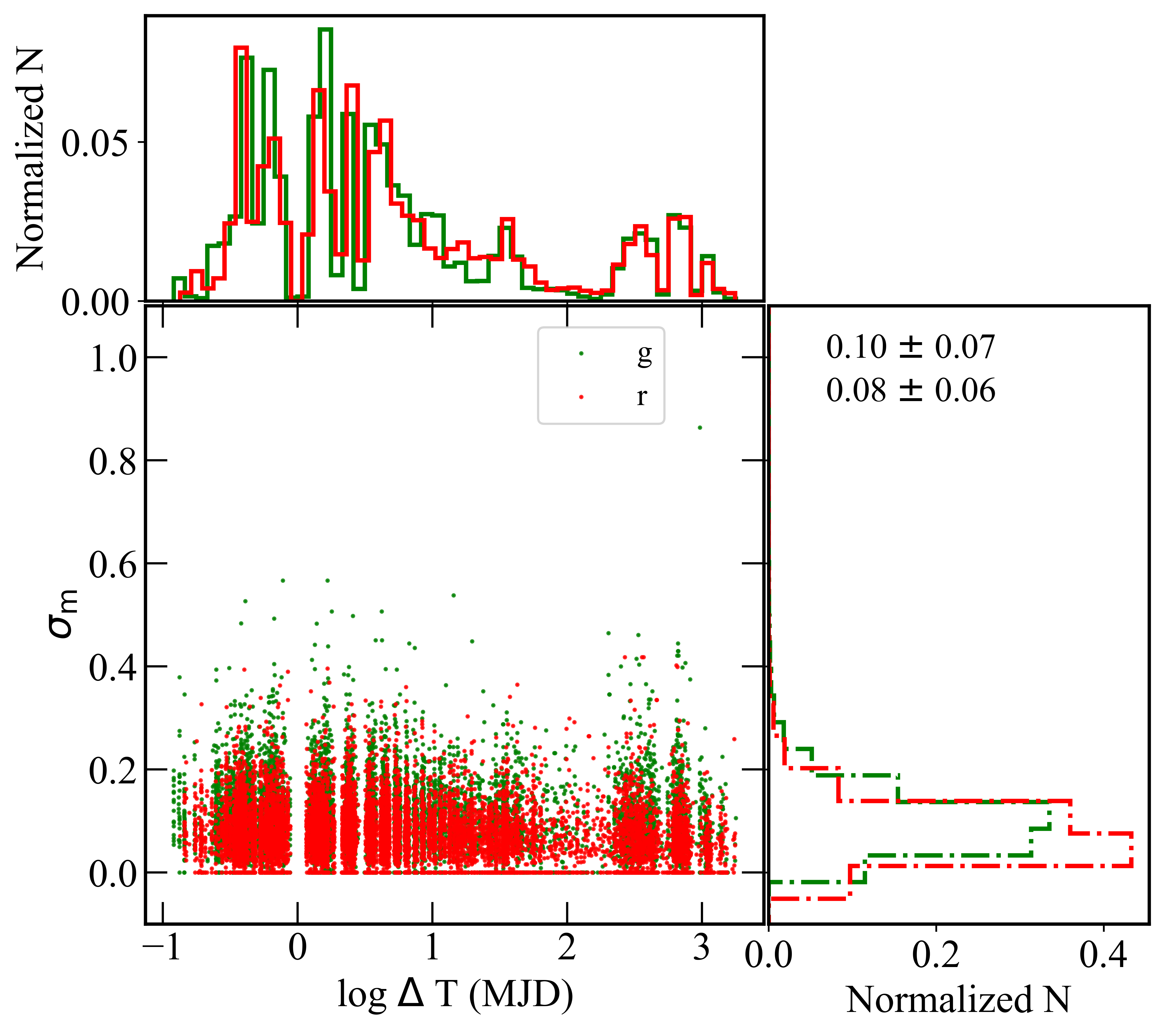}   
		\caption{The distribution of time interval between LAMOST and ZTF and the intrinsic variability of quasars in the ZTF $g$ and $r$ bands.}
\label{fig:inter_var} 
\end{figure}

LAMOST is a spectroscopic survey telescope that does not conduct its own photometric measurements. Its field of view typically lack a sufficient number of standard stars, limiting its flux calibration to a relative scale \citep[e.g.,][]{2015MNRAS.448...90X}. In previous works, the broadband photometric measurements from SDSS \citep{2000AJ....120.1579Y} or Pan-STARRS1 \citep{2016arXiv161205560C,2020ApJS..251....7F} were used to perform the absolute flux calibration \citep[][]{2019ApJS..240....6Y,2023ApJS..265...25J}. However, the significant time interval between photometric and spectroscopic observations, often spanning years, would introduces substantial uncertainties due to the intrinsic long-term variability of quasars. To overcome this limitation, we utilized (quasi-)simultaneous photometric data from ZTF to recalibrate the absolute flux. We cross-matched the LAMOST quasar catalog with the ZTF database \footnote{\url{https://www.ztf.caltech.edu/ztf-public-releases.html}} with a 2\arcsec\, matching radius. Due to the limited monitoring cadence of the ZTF-$i$ band data, we only used $g$ and $r$ bands data to recalibrate spectra from the blue and red channels, respectively.  { We present the distribution of time interval and intrinsic variability of quasars in \autoref{fig:inter_var}, where the intrinsic variability is estimated following the method described in \citet{2019MNRAS.483.2362R}. 5715 (8217) sources have the time separation less than 3 (10) days, and the mean intrinsic variability for $g$ and $r$ bands are $0.1\pm0.07$ and  $0.08\pm0.06$. ZTF's public survey operates on a 3-night cadence on average, providing contemporary photometry that minimizes variability-induced errors. }

\begin{figure*}[!htb]
\centering
\includegraphics[page=1,width=0.5\textwidth]{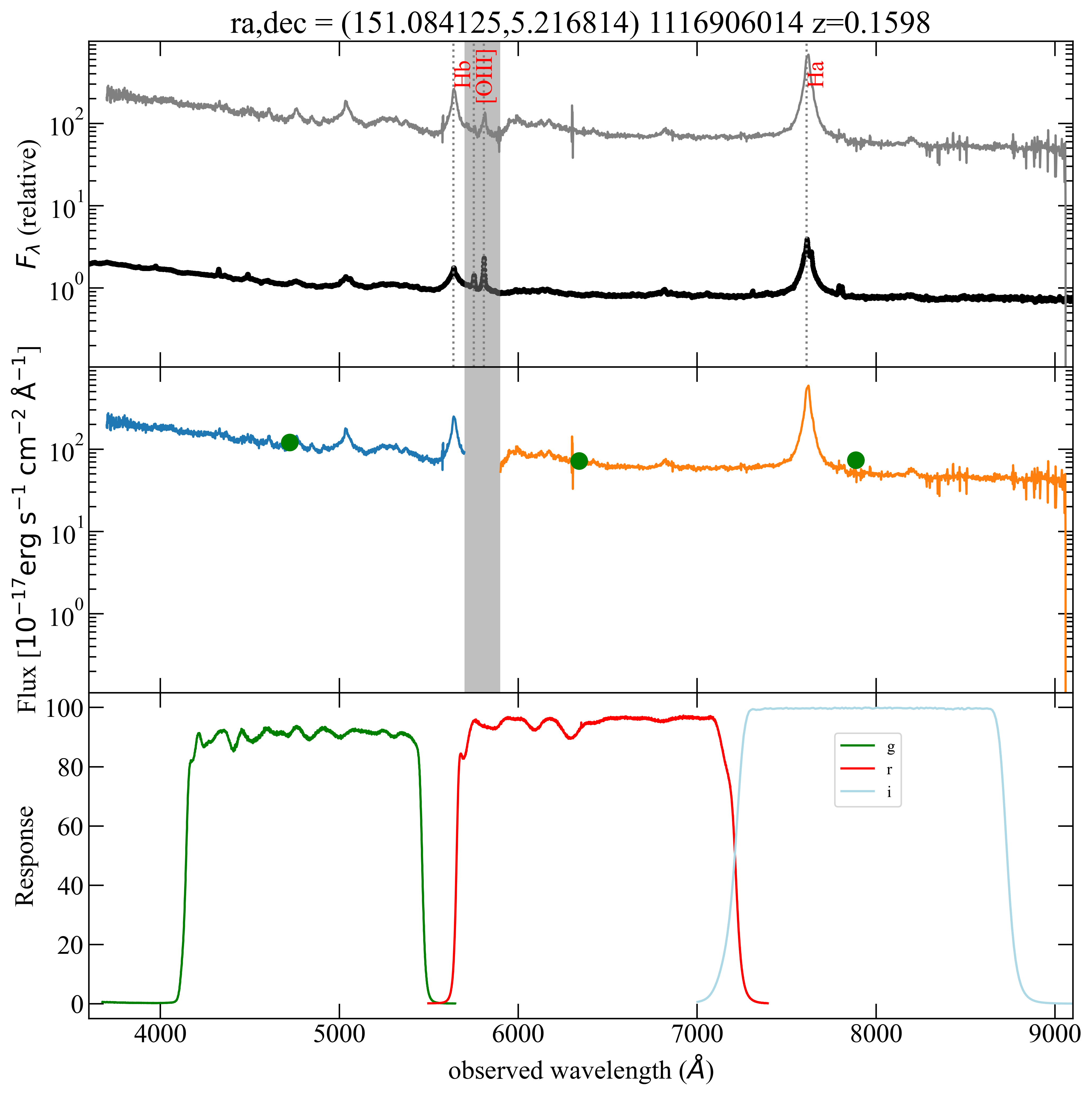}   
\caption{An example of absolute flux calibration for the blue-channel and red-channel LAMOST spectra. Top panel: The grey line represents the original spectrum only with the relative flux calibration. The black line represents the quasar template to match the spectrum. The gray area represents the blue-red overlapping region. Middle panel: The spectrum after the absolute flux calibration. The green dots represent the flux densities in the g, r, and i bands. The bottom panel shows the filter curves for the ZTF in the g, r, and i bands.}
\label{fig:RedBlue} 
\end{figure*}

The absolute flux calibration was performed as follows. We computed the synthetic photometry from the LAMOST spectra with the tool PYPHOT \footnote{\url{https://mfouesneau.github.io/pyphot/}}. We then calculated the flux scaling factors based on the differences between the magnitudes from LAMOST spectra and the photometric data from ZTF at the nearest epoch with LAMOST observation. In order to ensure robust and accurate measurements of flux, we applied \texttt{Lightkurve}, a Python package to remove the outlier data points in the light curves with 3$\sigma$ clip \citep{2018ascl.soft12013L} and required that the photometric uncertainties $magerr < 0.15$ mag \citep[e.g.,][]{2018ApJ...854..160R}. The scaling factors were applied to LAMOST spectra in the blue and red channels separately to obtain absolutely flux-calibrated spectra, while the original spectral shapes were preserved. \autoref{fig:RedBlue} presents an example of absolute flux calibration for LAMOST spectra. {The final calibrated LAMOST spectra are available at the Zenodo repository \dataset[doi:10.5281/zenodo.17719426]{https://doi.org/10.5281/zenodo.17719426}.} 

\subsection{Spectral fitting} \label{sec:fit}
We employ a spectral fitting methodology similar to \citet[][]{2023ApJS..265...25J}. Prior to fitting, the overlap region ($5700-5900 \AA$ in the observed frame) between red and blue channels is masked out. The absolutely flux-scaled spectra are corrected for Galactic extinction with the dust map of \cite{1998ApJ...500..525S} and the Milky Way extinction law from \cite{2019ApJ...877..116W} with $R_{V} = 3.1$. The dereddened spectra are then shifted to the rest frame using visually confirmed redshifts. The spectra are fitted by ({\tt QSOFITMORE}; \citealt{2021zndo...5810042F}), a wrapper package for the publicly available multicomponent spectral fitting code {\tt pyQSOFit} \citep{2018ascl.soft09008G}. The detailed description of the fitting procedure can be found in \cite{2018ascl.soft09008G}, \cite{2019ApJS..241...34S}, \citet{2021zndo...5810042F}, and \citet{2022ApJS..261...32F}.

The host galaxy contribution is only $\sim 15\%$ on average for low-z ($z \lesssim 0.5$ ) and low-luminosity quasars, which corresponds to a little bit ($\sim 0.06$ dex) overestimation of 5100 $\rm \AA$ luminosity  \citep{2011ApJS..194...45S}. The host decomposition for faint sources with poor signal-to-noise ratio (S/N) may lead to larger uncertainties. Besides, it is negligible for high-z ($z \gtrsim 0.5$) or high luminosity ($\rm log L_{5100} \gtrsim$ 44.5) quasars. Thus, we do not apply the decomposition of the host galaxy to the spectra.

\subsubsection{Continuum} \label{sec:continuum}
The pseudocontinuum is fitted by a broken power law ($f$$\rm_{bpl}$) and a Fe\,{\sc ii} model \citep[$f$$\rm_{Fe\,{\sc ii}}$;][]{2001ApJS..134....1V}.  The turning point of the broken power law is fixed at 4661 $\rm \AA$ at rest-frame \cite[e.g.,][]{2001AJ....122..549V}. The Fe\,{\sc ii} template consist of the optical Fe\,{\sc ii} template from  \cite[][]{1992ApJS...80..109B}, and UV Fe\,{\sc ii} templates in the wavelength of 1000-2000 $\rm \AA$ from \cite{2001ApJS..134....1V}, 2200-3090 $\rm \AA$ from \cite{2007ApJ...662..131S}, and 3090-3500 $\rm \AA$ from \cite{2006ApJ...650...57T}, respectively. The iron model $f$$\rm_{Fe~\textsc{ii}}$ is
\begin{equation}
f_{\rm Fe~\textsc{ii}} = b_{0}F{\rm_{Fe~\textsc{ii}}}(\lambda,b_{1},b_{2}),
\end{equation}
where the parameters $b_{0}$, $b_{1}$, $b_{2}$ are the normalization, the full width at half-maximum (FWHM) of Gaussian profile, and the wavelength shift applied to the Fe\,{\sc ii} template.

For a few spectra with peculiar shapes in the continuum that might be caused by unstable
efficiencies of some fibers and poor relative flux calibrations, we add a three-order polynomial model ($f$$\rm_{poly}$) to fit the continuum \citep{2020ApJS..249...17R,2022ApJS..261...32F}. The peculiar shape in the continuum of these LAMOST spectra is evident compared to SDSS spectra. Only $\lesssim 0.3 \%$ of objects require an additional polynomial component. The final pseudocontinuum consists of two (or three) components as follows:

\begin{equation}
f_{\rm cont} = f{\rm_{bpl}} + f{\rm_{Fe~\textsc{ii}}} + (f\rm_{poly}).
\end{equation}

The pseudocontinuum component is subtracted and the remaining emission line component is fitted with Gaussian profiles. We concentrate on the four typical broad emission lines (H$\alpha$, H$\beta$, Mg{\sc ii} and
C{\sc iv}), which are the strongest broad emission lines for quasars in the LAMOST spectra and can serve as virial black hole mass estimators.  The parameters we are mainly interested in are FWHM, equivalent width (EW), and flux of emission lines. The fitting procedures for each line are described below. 

\begin{figure*}[!htb]
\centering
		\includegraphics[page=1,width=0.75\textwidth]{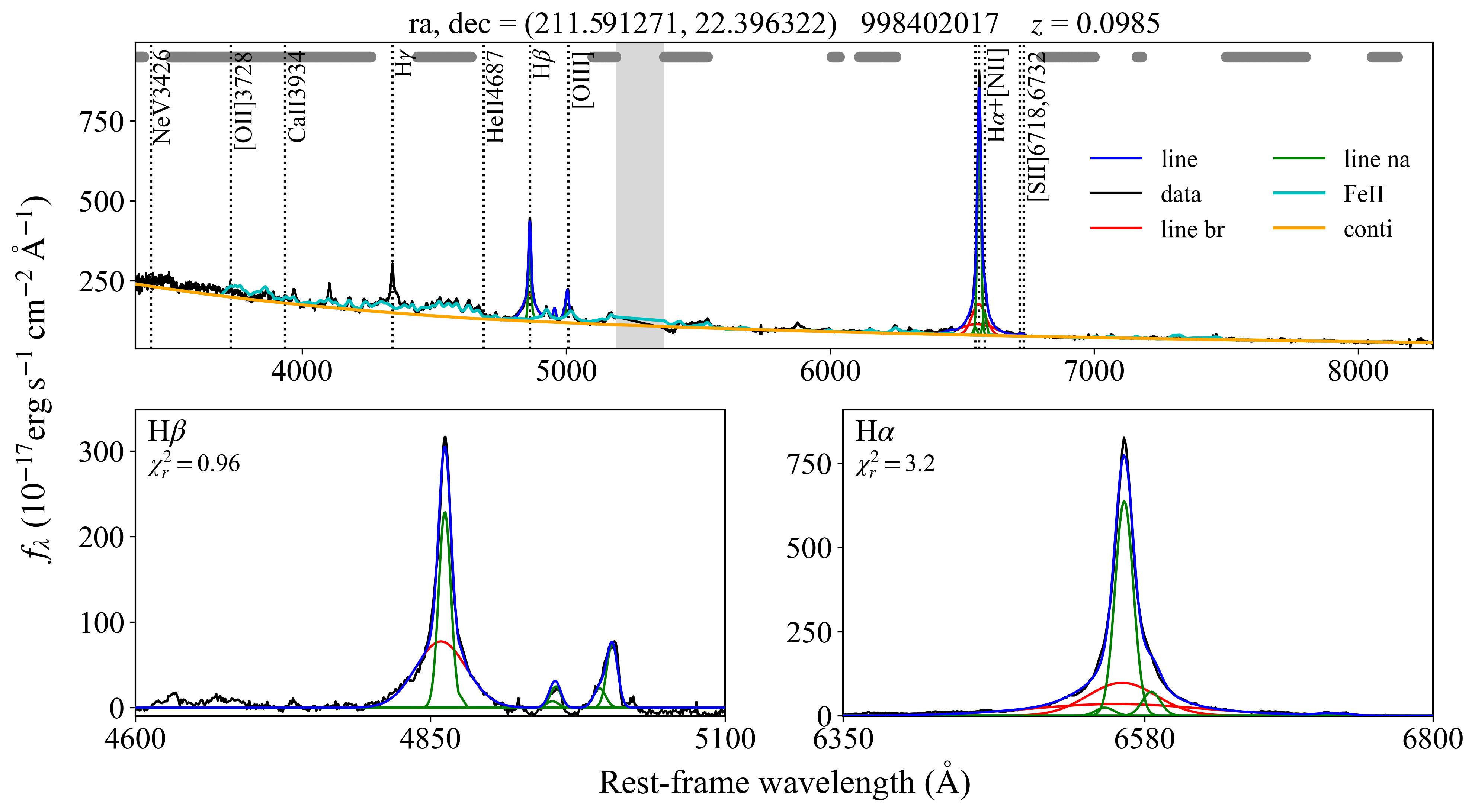}   
		\caption{An example for the spectral fitting for the LAMOST spectrum of a quasar with z=0.0985. The black lines represent the extinction-corrected spectra with the continuum subtracted in the lower panels. As for the fitted emission lines, the broad components are in red while the narrow ones are in green, along with their sum (blue). The H$\alpha$ and H$\beta$ emission lines are well fitted. } 
\label{fig:fit1} 
\end{figure*}

\begin{figure*}[!htb]
\centering
\includegraphics[page=1,width=0.75\textwidth]{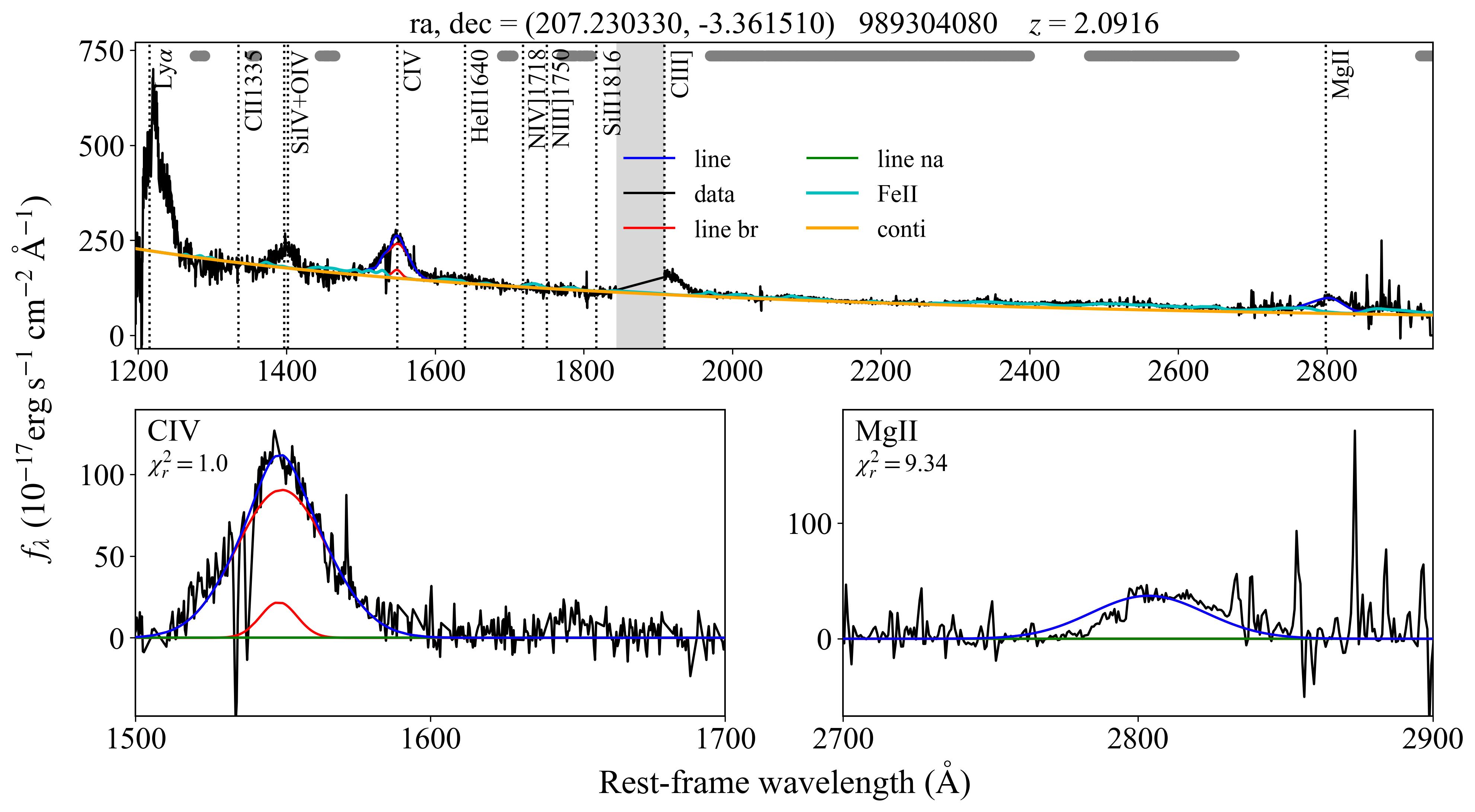}   
		\caption{Similar to \autoref{fig:fit1}, but for a quasar with z=2.0916. The Mg{\sc ii} and C{\sc iv} emission lines are well fitted.} 
\label{fig:fit4} 
\end{figure*}

\subsubsection{H$\alpha$ line}
The pseudocontinuum subtracted H$\alpha$-[N\,{\sc ii}]-[S\,{\sc ii}] emission lines are fitted in the rest-frame wavelength range [6350,6800] $\rm \AA$ for objects at $z \lesssim 0.37$. The broad component of H$\alpha$ is modeled by two Gaussian profiles, and the narrow components of H$\alpha$, [N\,{\sc ii}]$\lambda\lambda$6548,6585 and [S\,{\sc ii}]$\lambda\lambda$6718,6732 are each modeled by a single Gaussian profile. The upper limit of FWHM for the narrow components is set to be 900 km $\rm s^{-1}$, which is a commonly used FWHM criterion to separate the narrow and broad lines in quasars \citep{2009ApJ...707.1334W,2019A&A...625A.123C,2019ApJ...882....4W}. The line widths and velocity offsets of the narrow lines are tied to each other. The relative flux ratio of the [N\,{\sc ii}]$\lambda\lambda$6548,6585 doublet is fixed to 2.96.

\subsubsection{H$\beta$ line}
The pseudocontinuum subtracted H$\beta$-[O\,{\sc iii}] emission lines are fitted in the rest-frame wavelength range [4600,5100] $\rm \AA$ for objects at $z \lesssim 0.8$. The broad component of H$\beta$ is modeled by two Gaussian profiles, and the narrow component of H$\beta$ is modeled by a single Gaussian profile. The upper limit of FWHM for the narrow components is set to be 900 km $\rm s^{-1}$. In addition to the narrow component, the [O\,{\sc iii}]$\lambda\lambda$4959,5007 double lines are modeled with two Gaussians: one for the line core and another for the blue-shifted wing, and neither of them are tied to the H$\beta$ narrow component.  This approach is motivated by previous studies \citep{2005AJ....130..381B, 2004A&A...413.1087C, 2007ApJ...667L..33K, 2010MNRAS.403.1759Z, 2018A&A...615A..13S}. The line widths and velocity offsets of the cores and wings are tied to each other. We constrain the relative flux ratio of [O\,{\sc iii}]$\lambda\lambda$4959,5007 double lines to be the theoretical ratio of 1:3.

\subsubsection{Mg{\sc ii} line}
The Mg{\sc ii} and C{\sc iv} emission lines might be affected by the broad and narrow absorption features sometimes during the fitting process. In order to reduce the effect of narrow absorption features, we used ``$\rm rej\_abs = True$''  option in the {\tt QSOFITMORE} code when fitting Mg{\sc ii} and C{\sc iv} emission lines. The code masks out the 3$\sigma$ outliers below the continuum model, which is helpful to reduce the impact of absorption features \citep{2011ApJS..194...45S,2019ApJ...874...22S}.

The Mg{\sc ii} emission lines are fitted in the rest-frame wavelength range of [2700,2900] $\rm \AA$ for objects at 0.36 $ \lesssim z \lesssim$ 2.1. The broad component of Mg{\sc ii} is modeled by two Gaussian profiles. For the narrow component, some AGNs show the Mg{\sc ii}$\lambda\lambda$2796,2803 double lines around the peak, and the FWHM of each component is $\lesssim$ 750 km $\rm s^{-1}$ \citep{2011ApJS..194...45S}. However, such cases are rare and most LAMOST spectra do not have adequate S/N and/or spectral resolution to separate these two components. Additionally, the narrow Mg{\sc ii} absorption line can lead to a mimicked double-peak profile. Therefore, we fit the Mg{\sc ii} narrow component with a single narrow Gaussian with an FWHM upper limit of 900 km $\rm s^{-1}$. 

\subsubsection{C{\sc iv} line}
The C{\sc iv} emission lines are fitted in the spectral rest-frame wavelength range of [1500,1700]$\rm \AA$ for objects at 1.5 $ \lesssim z \lesssim$ 4.4 . Similar to other emission lines, the broad component of the C{\sc iv} line is modeled by two Gaussian profiles. It is still debatable whether a strong narrow C{\sc iv} component exists for most quasars \citep{2011ApJ...742...93A, 2012ApJ...759...44D, 2019ApJS..241...34S}. Thus, we do not set an upper limit for the FWHM of the narrow component. It is uncertain whether the narrow component subtraction is feasible for the C{\sc iv} emission line, and the existing  C{\sc iv} virial estimators are calculated with the FWHM from the entire C{\sc iv} profiles. The parameters of the entire C{\sc iv} profile are also provided in Table~\ref{tab:catalog2}.  Examples of the best-fitting results of H$\alpha$, H$\beta$, Mg{\sc ii}, and C{\sc iv} lines are given in \autoref{fig:fit1} and  \autoref{fig:fit4}.

\subsubsection{The Reliability of the Spectral Fitting and Error Estimation}
Following the automated spectral fitting procedure, we conducted a systematic visual inspection of the results for each spectrum. The fitting quality is generally acceptable for the majority of spectra with high S/N. Poor fittings are primarily due to low S/N and an insufficient number of good pixels in the fitting region. We assigned a flag to each line based on the visual inspection:
{\tt LINE\_FLAG = 0} indicates an acceptable fitting and reliable measurement; {\tt LINE\_FLAG = -1} indicates a spurious fitting; {\tt LINE\_FLAG = -9999} indicates that there are not enough good pixels in the fitting region, often due to limitations in spectral quality or wavelength coverage. Furthermore, the presence of broad absorption line (BAL) features can significantly affect the fitting. { BAL features present at Mg{\sc ii}, C{\sc iv}, and both lines are marked with {\tt BAL\_FLAG = 1,2,3}, respectively.}

\begin{figure}[!htb]
\includegraphics[page=1,width=0.4\textwidth]{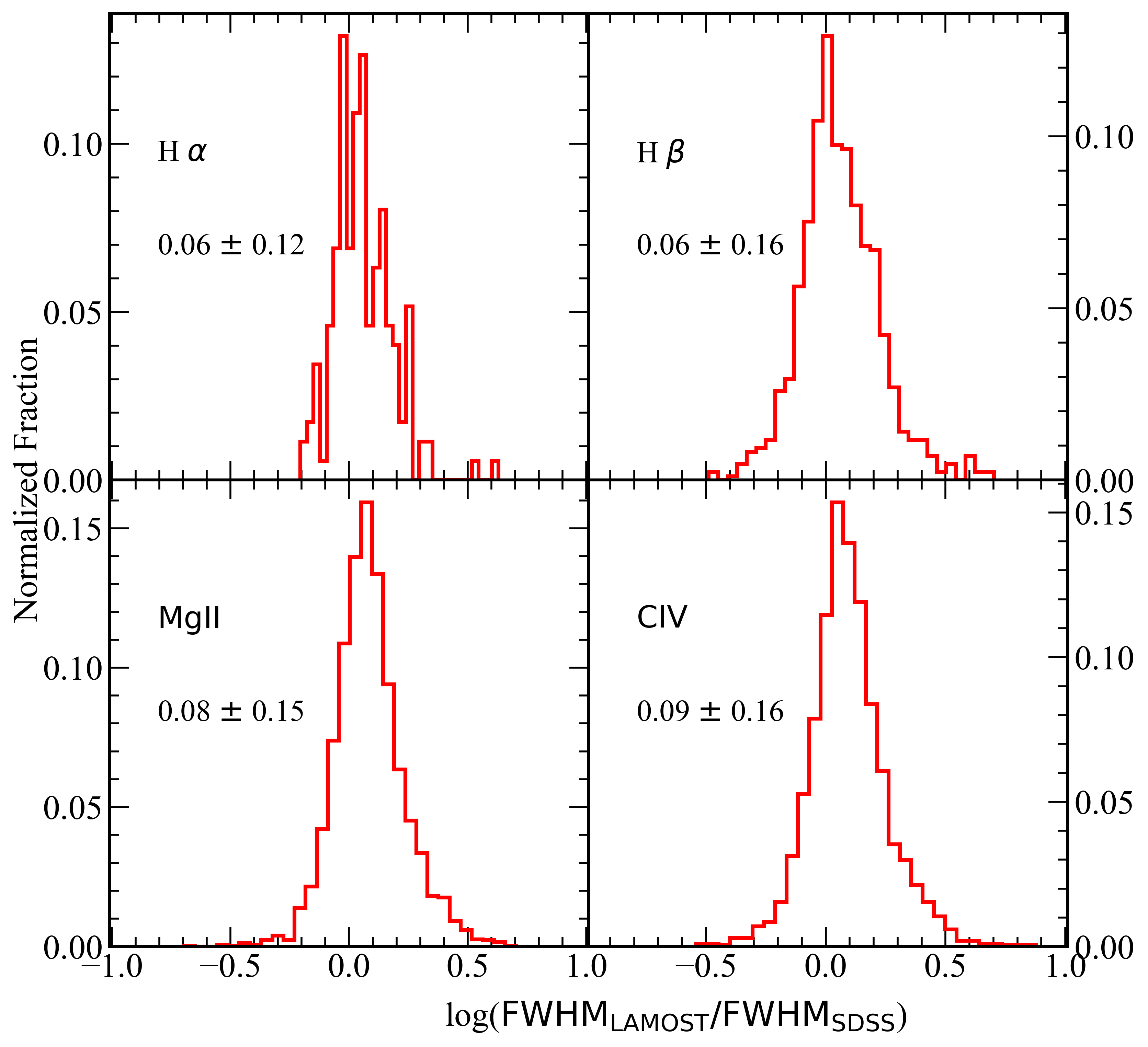}   
		\caption{Comparisons between the measurements of the FWHM values in this work and \citet{2022ApJS..263...42W}. We show the plot of $\rm log(FWHM_{LAMOST}/FWHM_{SDSS})$ for broad H$\alpha$ (upper left), broad H$\beta$ (upper right), broad Mg{\sc ii} (lower left) and whole C{\sc iv} (lower right). } 
\label{fig:Com_FWHM} 
\end{figure}

\begin{figure}[!htb]
\includegraphics[page=1,width=0.4\textwidth]{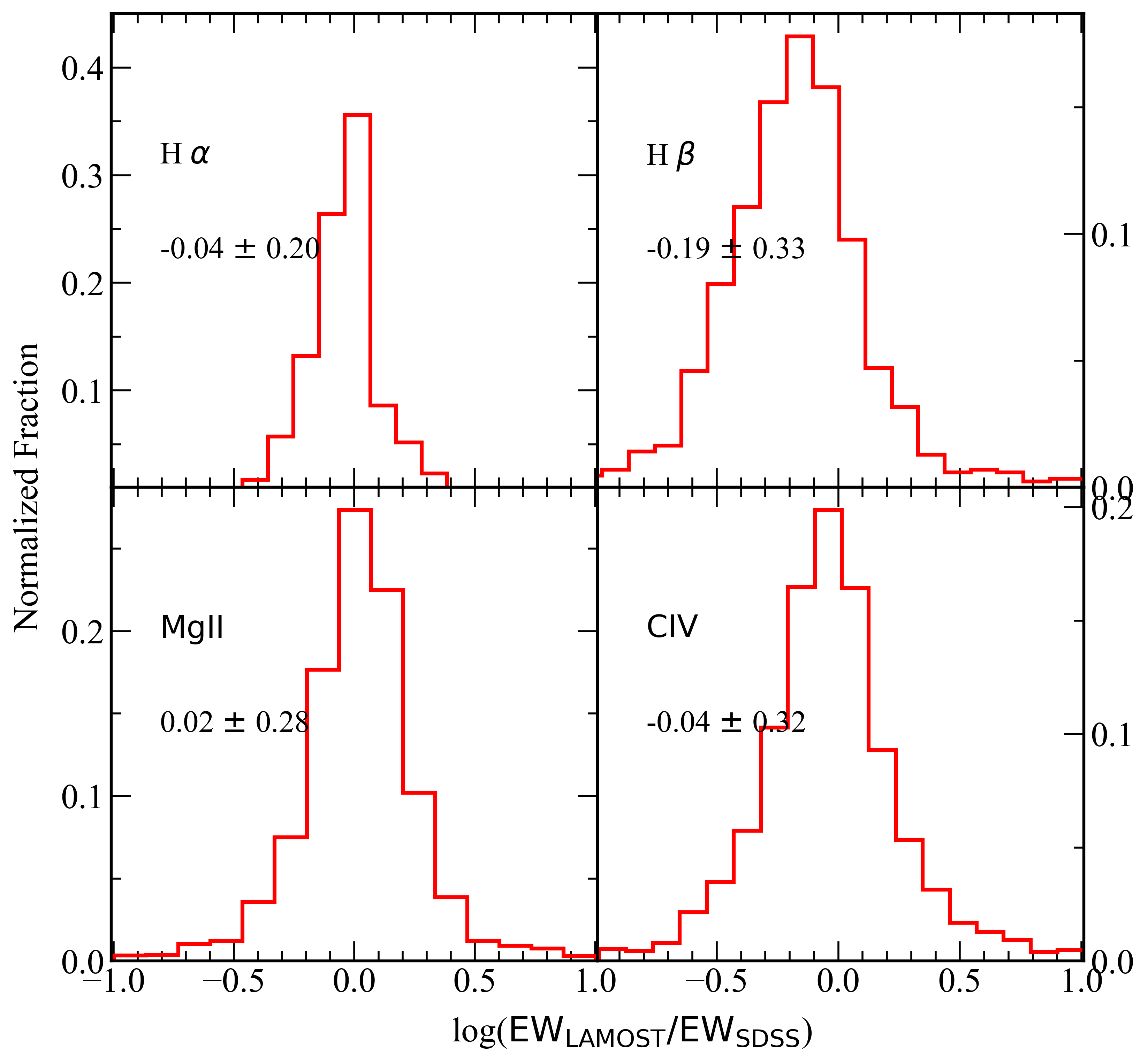}   
		\caption{Same as in Figure ~\ref{fig:Com_FWHM}, but for EW values.} 
\label{fig:Com_EW} 
\end{figure}

\begin{figure}[!htb]
\includegraphics[page=1,width=0.4\textwidth]{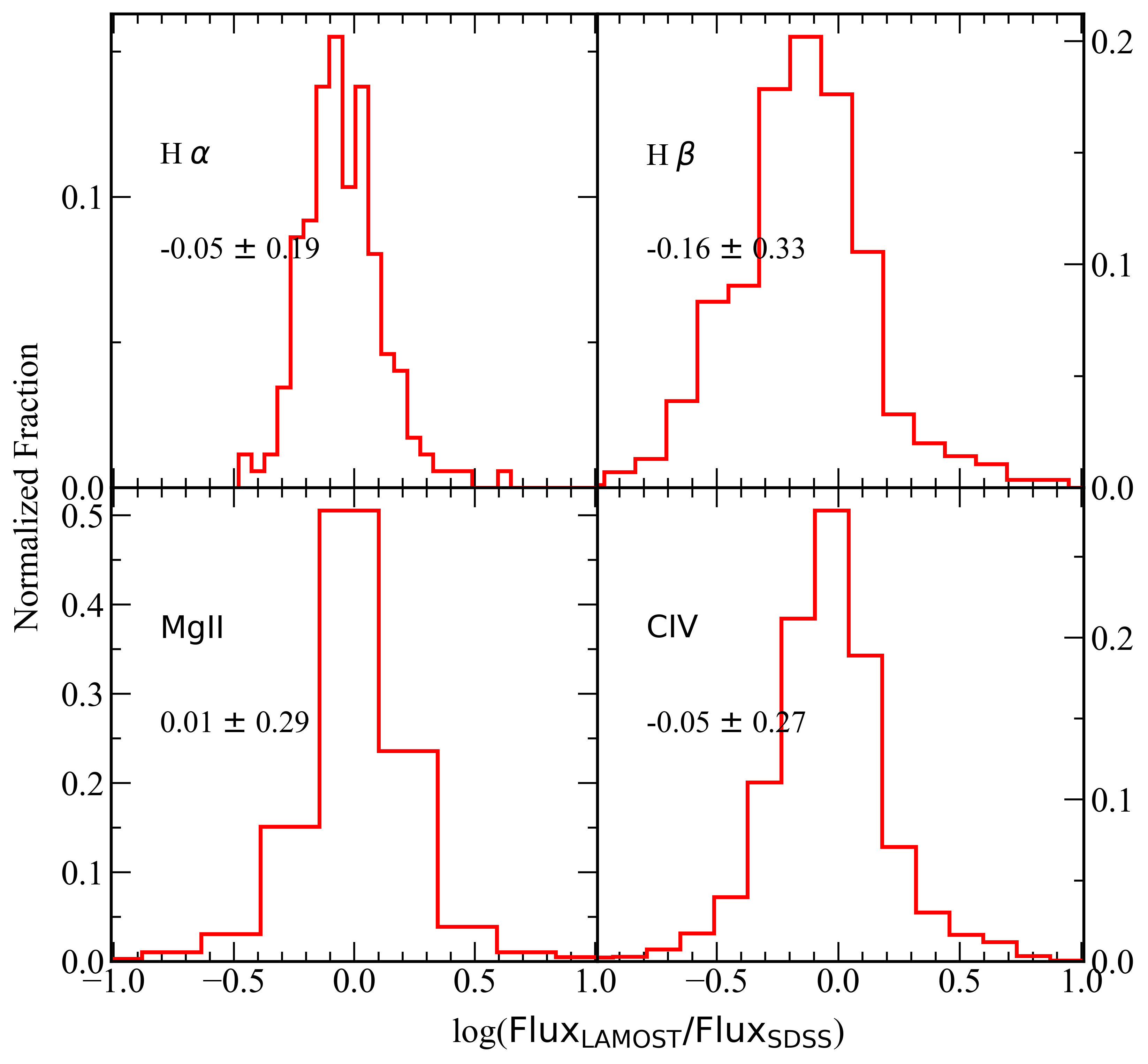}   
		\caption{Same as in Figure ~\ref{fig:Com_FWHM}, but for emission line flux.} 
\label{fig:Com_Flux} 
\end{figure}

\begin{figure}[!htb]
\includegraphics[page=1,width=0.4\textwidth]{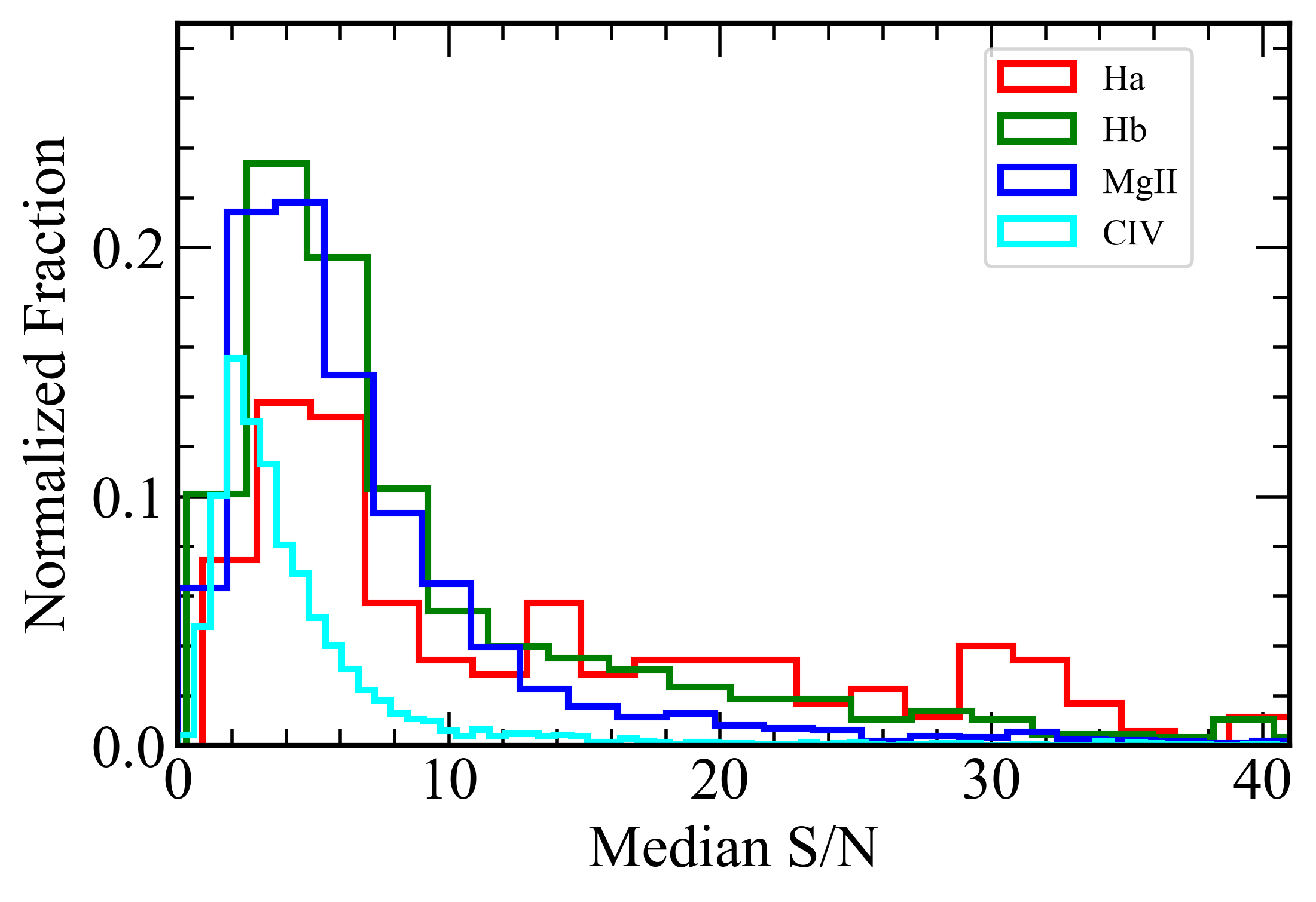}   
		\caption{The normalized distributions of median S/N per pixel around the line-fitting region. } 
\label{fig:Fraction} 
\end{figure}

Our catalog contains $5,057$ quasars in common with SDSS DR16Q quasar catalog \citep{2022ApJS..263...42W}. To validate our spectral fitting results, we further compare the measurements of key parameters for these overlapping sources. \autoref{fig:Com_FWHM},  \autoref{fig:Com_EW}, and  \autoref{fig:Com_Flux} show the comparison for FWHM, EW, and flux values, respectively. The parameters exhibit excellent overall agreement.  For the logarithmic FWHM values, the mean ($\mu$) and standard deviation ($\sigma$) of the difference between this work and \citet{2022ApJS..263...42W} are 0.06$\pm$ 0.12, 0.06$\pm$ 0.16, 0.08$\pm$ 0.15, and 0.09$\pm$ 0.16 for H$\alpha$, H$\beta$, Mg{\sc ii}, and C{\sc iv} emission lines, respectively.  The $\mu$ and $\sigma$ for the EW values of the difference between this work and \citet{2022ApJS..263...42W} are -0.04$\pm$ 0.20, -0.19$\pm$ 0.33, 0.02$\pm$ 0.28, and -0.04$\pm$ 0.32 for four typical emission lines. Similarly, the $\mu$ and $\sigma$ for the flux difference are -0.05$\pm$ 0.19, -0.16$\pm$ 0.33, 0.01$\pm$ 0.29, and -0.05$\pm$ 0.27. The slight discrepancy might be caused by the quasar variability and different spectral quality of LAMOST and SDSS. The peaks of median S/N per pixel around the line-fitting region are around 5 and even lower for LAMOST spectra as shown in \autoref{fig:Fraction}, which is significantly lower than that of SDSS spectra. Besides, different models are used in the spectral fitting. The continuum is fitted with a combination of a power law and a third-order polynomial in \citet{2022ApJS..263...42W}. For the emission-line-fitting process, three Gaussians are used to model the broad component of H$\alpha$ and H$\beta$ emission lines in \citet{2022ApJS..263...42W}, while we use two Gaussians. 

Uncertainties in spectral measurements are quantified via Monte Carlo (MC) approach. For each spectrum, we generated 20 mock spectra by adding Gaussian-distributed random noise at each pixel using the spectral-flux errors. The standard deviation of the resulting parameter distribution across all realizations defines the 1$\sigma$ uncertainty.

\subsection{Virial Black hole Mass}
Assuming that the broad line region (BLR) is virialized, the $\rm M_{BH}$ can be estimated from single-epoch spectrum based on the empirical scaling relation between monochromatic continuum luminosity and BLR radius as well as the relation between broad line width and virial velocity. We adopt the following empirical scaling relations to estimate the virial black hole masses,
\begin{equation}
\begin{split}
    \mathrm{l}&\mathrm{og}(M_{\mathrm{BH}}/M_{\odot}) = \\
    &\mathrm{log} \left[\left(\frac{\mathrm{FWHM}(\mathrm{H}\beta)}{\mathrm{km~s^{-1}}}\right)^2\left(\frac{L_{5100}}{10^{44}\mathrm{erg~s^{-1}}}\right)^{0.5}\right]+0.91
\end{split}
\end{equation}
for H$\beta$-based estimator from \citet{2006ApJ...641..689V}, 
\begin{equation}
\begin{split}
    \mathrm{l}&\mathrm{og}(M_{\mathrm{BH}}/M_{\odot}) =  \\
    &\mathrm{log} \left[\left(\frac{ \mathrm{FWHM (Mg{\sc II})} }{\mathrm{km~s^{-1}}}\right)^{1.51}\left(\frac{L_{3000}}{10^{44}\mathrm{erg~s^{-1}}}\right)^{0.5}\right]+2.60
\end{split}
\end{equation}
for Mg{\sc ii}-based estimator from \citet{2009ApJ...707.1334W}, and
\begin{equation}
\begin{split}
    \mathrm{l}&\mathrm{og}(M_{\mathrm{BH}}/M_{\odot}) = \\
    &\mathrm{log} \left[\left(\frac{ \mathrm{FWHM (C{\sc IV}) }}{\mathrm{km~s^{-1}}}\right)^{2}\left(\frac{L_{1350}}{10^{44}\mathrm{erg~s^{-1}}}\right)^{0.53}\right]+0.66
\end{split}
\end{equation}
for C{\sc iv}-based estimator from \citet{2006ApJ...641..689V}.

\begin{figure}[!htb]
\includegraphics[width=0.4\textwidth]{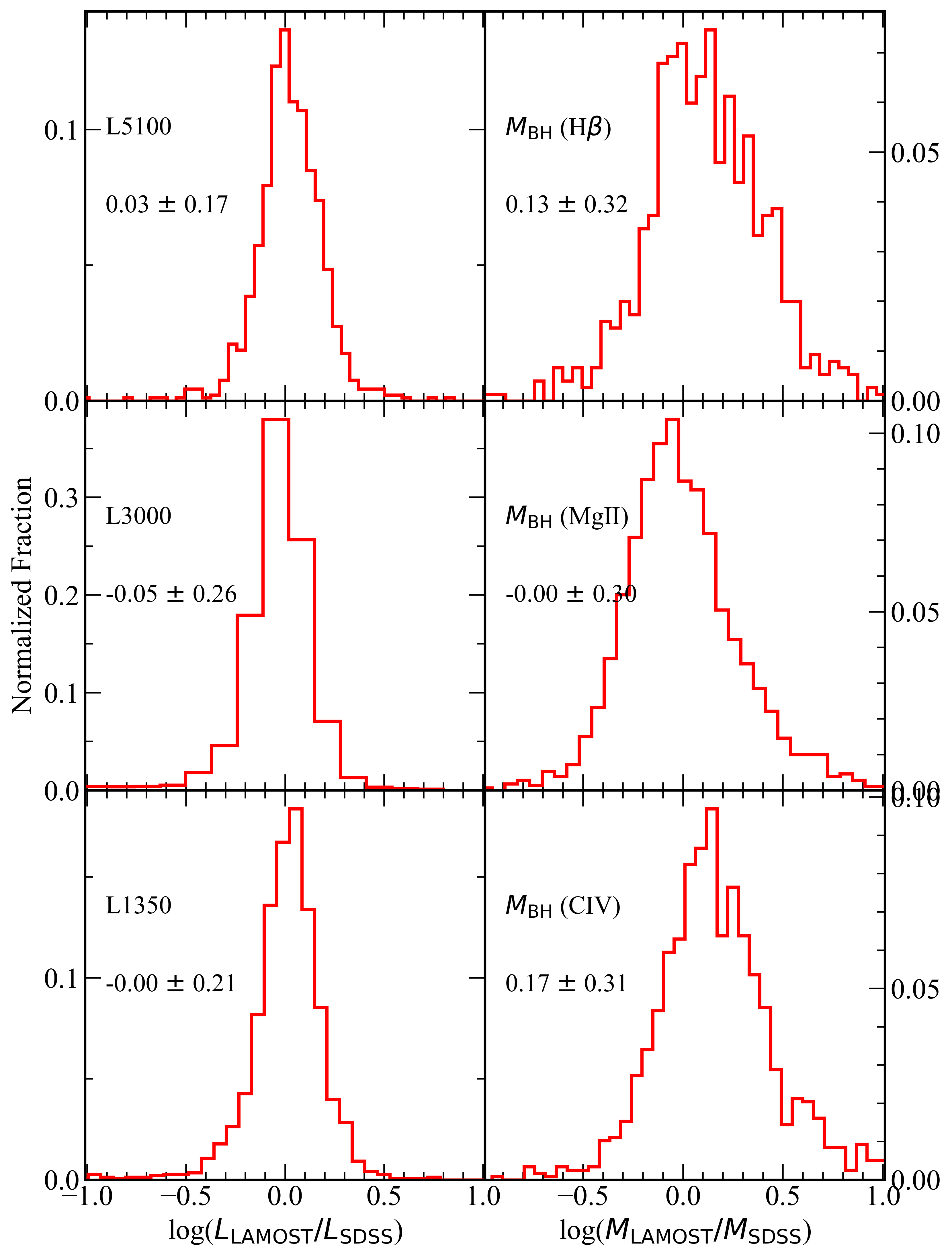 }   
		\caption{The comparison of the monochromatic continuum luminosities ($L_{5100}$, $L_{3000}$, $L_{1350}$) and the estimated $\rm M_{BH}$  based on H$\beta$, Mg{\sc ii} and C{\sc iv} between this work and \citet{2022ApJS..263...42W}.} 
\label{fig:Com_Lcom_BHmass} 
\end{figure}

The monochromatic continuum luminosities at 1350 ($L_{1350}$), 3000 ($L_{3000}$), and 5100 ($L_{5100}$) $\rm \AA$ are derived from the best-fit continuum of the absolutely flux-calibrated spectra. As illustrated in \autoref{fig:Com_Lcom_BHmass}, our measurements of monochromatic continuum luminosities show excellent agreement with those from \citet{2022ApJS..263...42W}. 

\begin{figure}[!htb]
\includegraphics[width=0.4\textwidth]{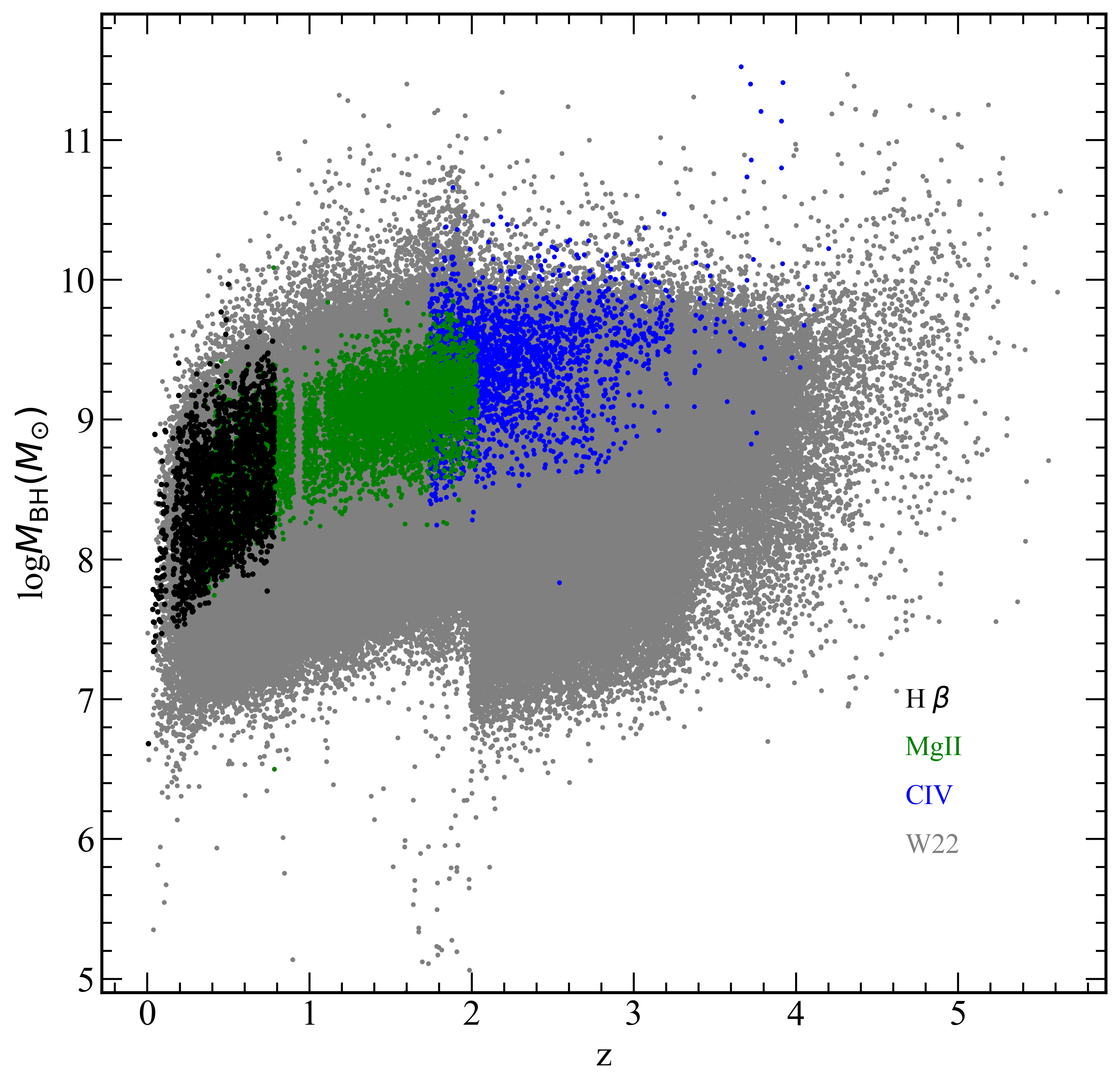}   
		\caption{The distribution of $\rm M_{BH}$ based on various broad emissions (H$\beta$, Mg{\sc ii} and C{\sc iv}) in this work is plotted against the redshift. The quasars from \citet[][]{2022ApJS..263...42W} are represented by the grey dots.} 
\label{fig:z_BHmass} 
\end{figure}

There is strong consistency for $\rm M_{BH}$ based on the Mg{\sc ii} and C{\sc iv} emission lines, but slight discrepancy for $\rm M_{BH}$ based on the H$\beta$ emission line. The distribution of $\rm M_{BH}$ as a function of redshift is presented in \autoref{fig:z_BHmass}. Notably, the LAMOST quasars occupy an intermediate parameter space in $\rm M_{BH}-z$ plane compared to SDSS DR16Q sample in \citet{2022ApJS..263...42W}. This intermediate distribution also indicates that our absolute flux calibration is robust and the $\rm M_{BH}$ measurements are reliable. 

\section{DESCRIPTION OF THE CATALOG} \label{sec:catalog}
We compile a catalog for the 11,346 quasars identified in LAMOST DR 10-12, which will be available online at LAMOST public website\footnote{https://nadc.china-vo.org/?locale=en}.
A summary of the parameters is listed in Table~\ref{tab:catalog2} and described below.

\begin{enumerate} \setlength{\itemsep}{-0.5ex}
	\item Unique spectra ID in LAMOST database. 
	\item Target Observation date. 
	\item LAMOST object designation: $\rm Jhhmmss.ss+ddmmss.s$ (J2000). \item Right Ascension and Declination (in decimal degrees, J2000). 
	\item Spectroscopic observation information: Local modified Julian date (LMJD), spectroscopic plan name (PlanID), spectrograph identification (spID), and spectroscopic fiber number (fiberID). These four numbers are unique for each spectrum named in the format of {spec$-$LMJD$-$planID\_spID$-$fiberID.fits}. 
	\item Redshift and its flag (ZWARNING) based on visual inspections. 1$=$not robust (eg., only one emission line available). \item Target selection flag. `SOURCE\_FLAG$=$1' indicates that the quasar was selected from sources classified as QSO by 1D pipeline. `SOURCE\_FLAG$=$0' means the object is a quasar candidate in the LAMOST input catalog . 
	\item $ {\rm M}_{i}$ ($z$=2): absolute i-band magnitude with K-corrected to $z=2$ following  \citet{2006AJ....131.2766R}. 
	\item Number of spectroscopic observations for the quasar. When there are more than one observations for the object, the line properties are obtained from only one of the observations in which the S/N is highest. 	
\item Median S/N per pixel in the continuum wavelength regions. 
	\item Flag of broad absorption features. BAL\_FLAG$=$1,2,3 indicate broad absorption features are present in Mg{\sc ii}, C{\sc iv}, and both lines, respectively. 
	\item FLUX, FWHM, rest-frame EW, and their uncertainties for broad \ha, narrow \ha\,, [N\,{\sc ii}]$\lambda6585$ and [S\,{\sc ii}]$\lambda\lambda$6718,6732 emission lines. 
	\item Number of good pixels and median S/N per pixel for the spectrum in \ha\ region of rest-frame 6350-6800\,\AA. 
	\item Flag indicates reliability of the emission line fitting results in \ha\ region upon visual inspections. 0$=$acceptable; -1$=$unacceptable; $-9999$ not measured due to too few good pixels in the fitting region. Flag also holds for other emission lines.
	\item FLUX, FWHM, rest-frame EW, and their uncertainties for broad \hb, narrow \hb, [O\,{\sc iii}] $\lambda 5007$ emission lines. 
	\item Number of good pixels and median S/N per pixel for the spectrum in \hb\ region of rest-frame 4600-5100\,\AA. 
	\item Flag indicates reliability of the emission line fitting results in \hb\ region upon visual inspections. 
	\item FLUX, FWHM, rest-frame EW, and their uncertainties for the broad and narrow \mgii\ emission line. 
	\item Number of good pixels and median S/N per pixel for the spectrum in \mgii\ region of rest-frame 2700-2900\,\AA. 
	\item Flag indicates reliability of the emission line fitting results in \mgii\ region upon visual inspections. 
	\item FLUX, FWHM, rest-frame EW, and their uncertainties for the whole, broad and narrow \civ\ emission line. 
	\item Number of good pixels and median S/N per pixel for the spectrum in \civ\ region of rest-frame 1500-1700\,\AA. 
	\item Flag indicates reliability of the emission line fitting results in \civ\ region upon visual inspections. 
	\item Wavelength power-law index, $\alpha_\lambda$, from blueward of 4661\,\AA. 
	\item Wavelength power-law index, $\alpha_\lambda$, from redward of 4661\,\AA. 
	\item Rest-frame normalization parameter of optical Fe\,{\sc ii}.
	\item Rest-frame Gaussian FWHM of optical Fe\,{\sc ii} complex.
	\item Rest-frame wavelength shift of optical Fe\,{\sc ii} complex.
	\item Rest-frame normalization parameter of UV Fe\,{\sc ii} complex.
	\item Rest-frame Gaussian FWHM of UV Fe\,{\sc ii}complex.
	\item Rest-frame wavelength shift of UV Fe\,{\sc ii} complex.
	\item Monochromatic luminosities and their uncertanties at 1350, 3000 and 5100\,\AA. 
	\item Virial black hole masses (in $\rm M_{\odot}$) with calibrations of \hb, \mgii\ and \civ. 
	\item Name of the quasar in SDSS quasar catalog. The LAMOST DR10-12 quasar catalog was cross-correlated with the SDSS quasar catalog \citep[DR16,][]{2022ApJS..263...42W} using a matching radius of 3\arcsec. \item Name of the object in second ROSAT all-sky survey point source catalog \citep[2RXS,][]{2016A&A...588A.103B}. The LAMOST DR10-12 quasar catalog was cross-correlated with 2RXS using a matching radius of 30\arcsec. 
	The nearest point source in 2RXS was chosen. 
	\item The background corrected source counts in full band (0.1-2.4\,keV), and its error, from 2RXS. 
	\item The exposure time of the ROSAT measurement. 
	\item Angular separation between the LAMOST and 2RXS source positions. 
	\item Name of the object in XMM-Newton Serendipitous Source catalog. The LAMOST DR10-12 quasar catalog was cross-correlated with XMM-Newton Serendipitous Source catalog \citep[4XMM-DR13,][]{2020A&A...641A.136W}  using a matching radius of 3\arcsec. \item The mean full-band (0.2-12\,keV) flux, and its error, from 4XMM-DR13. 
	\item Angular separation between the LAMOST and 4XMM-DR11 source positions. 
	\item FIRST peak flux density at 20\,cm in units of mJy. 
	The LAMOST DR10-12 quasar catalog was cross-correlated with FIRST survey catalog using a matching radius of 5\arcsec. 
	\item Angular separation between LAMOST and FIRST source positions. 
	\item SDSS (or Pan-STARRS1) $g$, $r$, $i$, $z$ PSF magnitudes without the correction for Galactic extinction, and their uncertainties.
	\item Flag of PSF magnitudes. {\tt MAG\_FLAG$=$1} indicates the PSF magnitudes are given by SDSS, {\tt MAG\_FLAG$=$0} indicates the PSF magnitudes are give by Pan-STARRS1 and {\tt MAG\_FLAG$=$-1} indicates that the quasars don‘t have reliable photometric information. 
	\item WISE $W1$, $W2$, $W3$ instrumental profile-fit photometry magnitudes without the correction for Galactic extinction, and their uncertainties.
	\item UKIDSS $Y$, $J$, $K$ AperMag3 magnitudes without the correction for Galactic extinction, and their uncertainties. The AperMag3 magnitudes are the aperture corrected magnitudes measured by UKIDSS with $\rm 2^{\prime \prime}$ diameter, providing the most accurate estimate of the total magnitude \citep{2006MNRAS.372.1227D}.
       \item  ZTF $g$ and $r$ band magnitude, uncertainties, and modified Julian date (MJD) for the observation used for absolute flux calibration and maximum variation magnitude.	
\end{enumerate}

\startlongtable
\begin{deluxetable*}{llll}
	\tabletypesize{\scriptsize} 
	\tablecaption{catalog format for the quasars identified in LAMOST DR10-12 \label{tab:catalog2}}
	\tablehead{\colhead{Column} &  \colhead{Name} &  \colhead{SI Units} &  \colhead{Description}}  
	\startdata
1   &      ObsID             & ---        &    Unique Spectra ID in LAMOST database  \\
2   &      ObsDate           & ---        &    Target observation date  \\
3   &      LAMOST            & ---        &    LAMOST designation hhmmss.ss+ddmmss (J2000)  \\
4   &      RAdeg             & deg        &    Right ascension (R.A.) in decimal degrees (J2000)  \\
5   &      DEdeg             & deg        &    Declination (Decl.) in decimal degrees (J2000)  \\
6   &      LMJD              & d          &    Local Modified Julian Day of observation  \\
7   &      PlanID            & ---        &    Spectroscopic plan identification  \\
8   &      spID              & ---        &    Spectrograph identification  \\
9   &      fiberID           & ---        &    Spectroscopic fiber number  \\
\hline
10  &      Z-VI              & ---        &    Redshift based on visual inspection  \\
11  &      ZWARNING          & ---        &    ZWARNING flag based on visual inspection  \\
12  &      SOURCE-FLAG       & ---        &    Flag of quasar candidate selection \\
13  &      Mimag-Z2          & mag        &    M$_{i} (z=2)$, K-corrected to $z=2$ following Richards et al. (2006)  \\
14  &      NSPECOBS          & ---        &    Number of spectroscopic observations  \\
15  &      SNR-SPEC          & ---        &    Median S/N per pixel of the spectrum  \\
16  &      BAL-FLAG          & ---        &    Flag of broad absorption features  \\
\hline
17  &      FLUX-Ha-b         & 10-20W.m-2 &    Flux of broad H{alpha} in $\rm 10^{-17} erg$ $\rm cm^{-2}$ $\rm s^{-1}$ \\
18  &   e\_FLUX-Ha-b         & 10-20W.m-2 &    Uncertainty in FLUX$_{\rm H\alpha,broad}$  \\
19  &      FWHM-Ha-b         & km.s-1     &    FWHM of broad H{alpha} in \kmps  \\
20  &   e\_FWHM-Ha-b         & km.s-1     &    Uncertainty in FWHM$_{\rm H\alpha,broad}$  \\
21  &      EW-Ha-b           & 10-10m     &    Rest-frame EW of broad H{alpha} in \AA  \\
22  &   e\_EW-Ha-b           & 10-10m     &    Uncertainty in EW$_{\rm H\alpha,broad}$  \\
23  &      FLUX-Ha-n         & 10-20W.m-2 &    Flux of narrow H{alpha} in $\rm 10^{-17} erg$ $\rm cm^{-2}$ $\rm s^{-1}$ \\
24  &   e\_FLUX-Ha-n         & 10-20W.m-2 &    Uncertainty in FLUX$_{\rm H\alpha,narrow}$  \\
25  &      FWHM-Ha-n         & km.s-1     &    FWHM of narrow H{alpha} in \kmps  \\
26  &   e\_FWHM-Ha-n         & km.s-1     &    Uncertainty in FWHM$_{\rm H\alpha,narrow}$  \\
27  &      EW-Ha-n           & 10-10m     &    Rest-frame EW of narrow H{alpha} in \AA  \\
28  &   e\_EW-Ha-n           & 10-10m     &    Uncertainty in EW$_{\rm H\alpha,narrow}$  \\
29  &      FLUX-NII-6585     & 10-20W.m-2 &    Flux of [N\,{\sc ii}]$\lambda$6585 in $\rm 10^{-17} erg$ $\rm cm^{-2}$ $\rm s^{-1}$ \\
30  &   e\_FLUX-NII-6585     & 10-20W.m-2 &    Uncertainty in FLUX$_{\rm[NII]6585}$  \\
31  &      FWHM-NII-6585     & km.s-1     &    FWHM of [N\,{\sc ii}]$\lambda$6585 in \kmps  \\
32  &   e\_FWHM-NII-6585     & km.s-1     &    Uncertainty in FWHM$_{\rm[NII]6585}$  \\
33  &      EW-NII-6585       & 10-10m     &    Rest-frame EW of [N\,{\sc ii}]$\lambda$6585 in \AA  \\
34  &   e\_EW-NII-6585       & 10-10m     &    Uncertainty in EW $_{\rm[NII]6585}$  \\
35  &      FLUX-SII-6718     & 10-20W.m-2 &    Flux of [S\,{\sc ii}]$\lambda$6718 in $\rm 10^{-17} erg$ $\rm cm^{-2}$ $\rm s^{-1}$ \\
36  &   e\_FLUX-SII-6718     & 10-20W.m-2 &    Uncertainty in FLUX$_{\rm[SII]6718}$  \\
37  &      FWHM-SII-6718     & km.s-1     &    FWHM of [S\,{\sc ii}]$\lambda$6718 in \kmps  \\
38  &   e\_FWHM-SII-6718     & km.s-1     &    Uncertainty in FWHM$_{\rm[SII]6718}$  \\
39  &      EW-SII-6718       & 10-10m     &    Rest-frame EW of [S\,{\sc ii}]$\lambda$6718 in \AA  \\
40  &   e\_EW-SII-6718       & 10-10m     &    Uncertainty in EW $_{\rm[SII]6718}$  \\
41  &      FLUX-SII-6732     & 10-20W.m-2 &    Flux of [S\,{\sc ii}]$\lambda$6732 in $\rm 10^{-17} erg$ $\rm cm^{-2}$ $\rm s^{-1}$ \\
42  &   e\_FLUX-SII-6732     & 10-20W.m-2 &    Uncertainty in FLUX$_{\rm[SII]6732}$  \\
43  &      FWHM-SII-6732     & km.s-1     &    FWHM of [S\,{\sc ii}]$\lambda$6732 in \kmps  \\
44  &   e\_FWHM-SII-6732     & km.s-1     &    Uncertainty in FWHM$_{\rm[SII]6732}$  \\
45  &      EW-SII-6732       & 10-10m     &    Rest-frame EW of [S\,{\sc ii}]$\lambda$6732 in \AA  \\
46  &   e\_EW-SII-6732       & 10-10m     &    Uncertainty in EW$_{\rm[SII]6732}$  \\
47  &      LINE-NPIX-HA      & ---        &    Number of good pixels for the rest-frame 6350-6800\,\AA  \\
48  &      LINE-MED-SN-HA    & ---        &    Median S/N per pixel for the rest-frame 6350-6800\,\AA  \\
49  &      LINE-FLAG-HA      & ---        &    Flag for the quality in H{alpha} fitting  \\
\hline
50  &      FLUX-Hb-b         & 10-20W.m-2 &    Flux of broad \hb\ in $\rm 10^{-17} erg$ $\rm cm^{-2}$ $\rm s^{-1}$ \\
51  &   e\_FLUX-Hb-b         & 10-20W.m-2 &    Uncertainty in FLUX$_{\rm H\beta,broad}$  \\
52  &      FWHM-Hb-b         & km.s-1     &    FWHM of broad \hb\ in \kmps  \\
53  &   e\_FWHM-Hb-b         & km.s-1     &    Uncertainty in FWHM$_{\rm H\beta,broad}$  \\
54  &      EW-Hb-b           & 10-10m     &    Rest-frame EW of broad \hb\ in \AA  \\
55  &   e\_EW-Hb-b           & 10-10m     &    Uncertainty in EW$_{\rm H\beta,broad}$  \\
56  &      FLUX-Hb-n         & 10-20W.m-2 &    Flux of narrow \hb\ in $\rm 10^{-17} erg$ $\rm cm^{-2}$ $\rm s^{-1}$ \\
57  &   e\_FLUX-Hb-n         & 10-20W.m-2 &    Uncertainty in FLUX$_{\rm H\beta,narrow}$  \\
58  &      FWHM-Hb-n         & km.s-1     &    FWHM of narrow \hb\ in \kmps  \\
59  &   e\_FWHM-Hb-n         & km.s-1     &    Uncertainty in FWHM$_{\rm H\beta,narrow}$  \\
60  &      EW-Hb-n           & 10-10m     &    Rest-frame EW of narrow \hb\ in \AA  \\
61  &   e\_EW-Hb-n           & 10-10m     &    Uncertainty in EW$_{\rm H\beta,narrow}$  \\
62  &      FLUX-OIII-5007    & 10-20W.m-2 &    Flux of [O\,{\sc iii}]$\lambda$5007 in $\rm 10^{-17} erg$ $\rm cm^{-2}$ $\rm s^{-1}$ \\
63  &   e\_FLUX-OIII-5007    & 10-20W.m-2 &    Uncertainty in FLUX$_{\rm[OIII]5007}$  \\
64  &      FWHM-OIII-5007    & km.s-1     &    FWHM of [O\,{\sc iii}]$\lambda$5007 in \kmps  \\
65  &   e\_FWHM-OIII-5007    & km.s-1     &    Uncertainty in FWHM$_{\rm[OIII]5007}$  \\
66  &      EW-OIII-5007      & 10-10m     &    Rest-frame EW of [O\,{\sc iii}]$\lambda$5007 in \AA  \\
67  &   e\_EW-OIII-5007      & 10-10m     &    Uncertainty in EW$_{\rm[OIII]5007}$  \\
68  &      LINE-NPIX-HB      & ---        &    Number of good pixels for the rest-frame 4600-5100\,\AA  \\
69  &      LINE-MED-SN-HB    & ---        &    Median S/N per pixel for the rest-frame 4600-5100\,\AA  \\
70  &      LINE-FLAG-HB      & ---        &    Flag for the quality in \hb\ fitting \\
\hline
71  &      FLUX-MgII-b       & 10-20W.m-2 &    Flux of the broad \mgii\ in $\rm 10^{-17} erg$ $\rm cm^{-2}$ $\rm s^{-1}$ \\
72  &   e\_FLUX-MgII-b       & 10-20W.m-2 &    Uncertainty in FLUX$_{\rm MgII,broad}$  \\
73  &      FWHM-MgII-b       & km.s-1     &    FWHM of the broad \mgii\ in \kmps  \\
74  &   e\_FWHM-MgII-b       & km.s-1     &    Uncertainty in FWHM$_{\rm MgII,broad}$  \\
75  &      EW-MgII-b         & 10-10m     &    Rest-frame EW of the broad \mgii\ in \AA  \\
76  &   e\_EW-MgII-b         & 10-10m     &    Uncertainty in EW$_{\rm MgII,broad}$  \\
77  &      FLUX-MgII-n       & 10-20W.m-2 &    Flux of the narrow \mgii\ in $\rm 10^{-17} erg$ $\rm cm^{-2}$ $\rm s^{-1}$ \\
78  &   e\_FLUX-MgII-n       & 10-20W.m-2 &    Uncertainty in FLUX$_{\rm MgII,narrow}$  \\
79  &      FWHM-MgII-n       & km.s-1     &    FWHM of the narrow \mgii$\lambda$ in \kmps  \\
80  &   e\_FWHM-MgII-n       & km.s-1     &    Uncertainty in FWHM$_{\rm MgII,narrow}$  \\
81  &      EW-MgII-n         & 10-10m     &    Rest-frame EW of the narrow \mgii\ in \AA  \\
82  &   e\_EW-MgII-n         & 10-10m     &    Uncertainty in EW$_{\rm MgII,narrow}$  \\
83  &      LINE-NPIX-MgII    & ---        &    Number of good pixels for the rest-frame 2700-2900\,\AA  \\
84  &      LINE-MED-SN-MgII  & ---        &    Median S/N per pixel for the rest-frame 2700-2900\,\AA  \\
85  &      LINE-FLAG-MgII    & ---        &    Flag for the quality in MgII fitting  \\
\hline
86  &      FLUX-CIV          & 10-20W.m-2 &    Flux of the whole \civ\ in $\rm 10^{-17} erg$ $\rm cm^{-2}$ $\rm s^{-1}$ \\
87  &   e\_FLUX-CIV          & 10-20W.m-2 &    Uncertainty in Flux$_{\rm CIV,whole}$  \\
88  &      FWHM-CIV          & km.s-1     &    FWHM of the whole \civ\ in \kmps  \\
89  &   e\_FWHM-CIV          & km.s-1     &    Uncertainty in FWHM$_{\rm CIV,whole}$  \\
90  &      EW-CIV            & 10-10m     &    Rest-frame EW of the whole \civ\ in \AA  \\
91  &   e\_EW-CIV            & 10-10m     &    Uncertainty in EW$_{\rm CIV,whole}$  \\
92  &      FLUX-CIV-b        & 10-20W.m-2 &    Flux of the broad \civ\ in $\rm 10^{-17} erg$ $\rm cm^{-2}$ $\rm s^{-1}$ \\
93  &   e\_FLUX-CIV-b        & 10-20W.m-2 &    Uncertainty in Flux$_{\rm CIV,broad}$  \\
94  &      FWHM-CIV-b        & km.s-1     &    FWHM of the broad \civ\ in \kmps  \\
95  &   e\_FWHM-CIV-b        & km.s-1     &    Uncertainty in FWHM$_{\rm CIV,broad}$  \\
96  &      EW-CIV-b          & 10-10m     &    Rest-frame EW of the broad \civ\ in \AA  \\
97  &   e\_EW-CIV-b          & 10-10m     &    Uncertainty in EW$_{\rm CIV,broad}$  \\
98  &      FLUX-CIV-n        & 10-20W.m-2 &    Flux of the narrow \civ\ in $\rm 10^{-17} erg$ $\rm cm^{-2}$ $\rm s^{-1}$ \\
99  &   e\_FLUX-CIV-n        & 10-20W.m-2 &    Uncertainty in Flux$_{\rm CIV,narrow}$  \\
100 &      FWHM-CIV-n        & km.s-1     &    FWHM of the narrow \civ\ in \kmps  \\
101 &   e\_FWHM-CIV-n        & km.s-1     &    Uncertainty in FWHM$_{\rm CIV,narrow}$  \\
102 &      EW-CIV-n          & 10-10m     &    Rest-frame EW of the narrow \civ\ in \AA  \\
103 &   e\_EW-CIV-n          & 10-10m     &    Uncertainty in EW$_{\rm CIV,narrow}$  \\
104 &      LINE-NPIX-CIV     & ---        &    Number of good pixels for the rest-frame 1500-1700\,\AA  \\
105 &      LINE-MED-SN-CIV   & ---        &    Median S/N per pixel for the rest-frame 1500-1700\,\AA  \\
106 &      LINE-FLAG-CIV     & ---        &    Flag for the quality in CIV fitting  \\
\hline
107 &      ALPHA-LAMBDA-1    & ---        &    Wavelength power-law index from blueward of 4661\,\AA  \\
108 &      ALPHA-LAMBDA-2    & ---        &    Wavelength power-law index from redward of 4661\,\AA  \\
109 &      Fe-op-norm        & ---        &    The normalization applied to the optical Fe\,{\sc ii} template  \\
110 &      Fe-op-shift       & ---        &    The Gaussian FWHM applied to the optical the Fe\,{\sc ii} template  \\
111 &      Fe-op-FWHM        & ---        &    The wavelength shift applied to the optical Fe\,{\sc ii} template  \\
112 &      Fe-uv-norm        & ---        &    The normalization applied to the ultraviolet Fe\,{\sc ii} template  \\
113 &      Fe-uv-shift       & ---        &    The Gaussian FWHM applied to the ultraviolet the Fe\,{\sc ii} template  \\
114 &      Fe-uv-FWHM        & ---        &    The wavelength shift applied to the ultraviolet Fe\,{\sc ii} template  \\
115 &      LOGL1350          & [10-7W]    &    Monochromatic luminosity at 1350\,\AA\ in $\rm erg\,s^{-1}$  \\
116 &   e\_LOGL1350          & [10-7W]    &    Uncertainty in $\rm log L_{1350}$ \\
117 &      LOGL3000          & [10-7W]    &    Monochromatic luminosity at 3000\,\AA\ in $\rm erg\,s^{-1}$  \\
118 &   e\_LOGL3000          & [10-7W]    &    Uncertainty in $\rm log L_{3000}$ \\
119 &      LOGL5100          & [10-7W]    &    Monochromatic luminosity at 5100\,\AA\ in $\rm erg\,s^{-1}$  \\
120 &   e\_LOGL5100          & [10-7W]    &    Uncertainty in $\rm log L_{5100}$ \\
\hline
121 &      LOGBH-HB          & [Msun]     &    Virial BH mass (M$_{\sun}$) based on \hb  \\
122 &      LOGBH-MgII        & [Msun]     &    Virial BH mass (M$_{\sun}$) based on \mgii  \\
123 &      LOGBH-CIV         & [Msun]     &    Virial BH mass (M$_{\sun}$) based on \civ  \\
124 &      SDSS-NAME         & ---        &    Name of the quasar in the SDSS quasar catalog  \\
125 &      2RXS-NAME         & ---        &    Name of the object in the 2nd ROSAT all-sky survey point source catalog  \\
126 &      2RXS-CTS          & ct         &    Background corrected source counts in 0.1-2.4\,keV from 2RXS source catalog  \\
127 &   e\_2RXS-CTS          & ct         &    Error of the source counts from 2RXS source catalog  \\
128 &      2RXS-EXPTIME      & s          &    Source exposure time from 2RXS source catalog  \\
129 &      LM-2RXS-SEP       & arcsec     &    LAMOST-2RXS separation in arcsec  \\
130 &      4XMM-NAME         & ---        &    Name of the object in XMM-Newton Serendipitous Source catalog  \\
131 &      4XMM-FLUX         & mW.m-2     &    Flux in 0.2-12.0\,keV band from 4XMM-DR11 (in erg\,s$^{-1}$\,cm$^{-2}$)  \\
132 &   e\_4XMM-FLUX         & mW.m-2     &    Error of the flux in 0.2-12.0\,keV band from 4XMM-DR11 (in erg\,s$^{-1}$\,cm$^{-2}$)  \\
133 &      LM-4XMM-SEP       & arcsec     &    LAMOST-4XMM separation in arcsec  \\
134 &      FPEAK             & mJy        &    FIRST peak flux density at 20 cm in mJy  \\
135 &      LM-FIRST-SEP      & arcsec     &    LAMOST-FIRST separation in arcsec  \\
\hline
136 &      gmag              & mag        &    SDSS (or Pan-STARRS1 PSF) g magnitudes  \\
137 &   e\_gmag              & mag        &    g PSF magnitude errors \\
138 &      rmag              & mag        &    SDSS (or Pan-STARRS1 ) r PSF magnitudes \\
139 &   e\_rmag              & mag        &    r PSF magnitude errors  \\
140 &      imag              & mag        &    SDSS (or Pan-STARRS1 ) i PSF magnitudes \\
141 &   e\_imag              & mag        &    i PSF magnitude errors \\
142 &      zmag              & mag        &    SDSS (or Pan-STARRS1 ) z PSF magnitudes \\
143 &   e\_zmag              & mag        &    z PSF magnitude errors \\
144 &      MAG-FLAG          & mag        &    Flag of PSF magnitude  \\
145 &      W1mag             & mag        &    instrumental profile-fit photometry magnitudes, W1 band \\
146 &   e\_W1mag             & mag        &    W1 magnitude errors \\
147 &      W2mag             & mag        &    instrumental profile-fit photometry magnitudes, W2 band \\
148 &   e\_W2mag             & mag        &    W2 magnitude errors \\
149 &      W3mag             & mag        &    instrumental profile-fit photometry magnitudes, W3 band \\
150 &   e\_W3mag             & mag        &    W3 magnitude errors  \\
151 &      Ymag              & mag        &    Y AperMag3 magnitudes ($\rm 2^{\prime \prime}$ aperture diameter) \\
152 &   e\_Ymag              & mag        &    Y magnitude errors \\
153 &      Jmag              & mag        &    J AperMag3 magnitudes ($\rm 2^{\prime \prime}$ aperture diameter) \\
154 &   e\_Jmag              & mag        &    J magnitude errors  \\
155 &      Kmag              & mag        &    K AperMag3 magnitudes ($\rm 2^{\prime \prime}$ aperture diameter) \\
156 &   e\_Kmag              & mag        &    K magnitude errors  \\
\hline	
157 &      zgmag             & mag        &    ZTF $g$ magnitudes used to calibrate flux  \\
158 &   e\_zgmag             & mag        &    ZTF $g$ magnitude errors \\
159 &      zrmag             & mag        &    ZTF $r$ magnitudes \\
160 &   e\_zrmag             & mag        &    ZTF $r$ magnitude errors  \\
161 &      MJD-zg            & d          &    MJD for ZTF $g$ band observation  \\
162 &      MJD-zr            & d          &    MJD for ZTF $r$ band observation  \\
163 &      Delta-zg          & mag        &    maximum magnitude variation for ZTF $g$ band observation  \\
164 &      Delta-zr          & mag        &    maximum magnitude variation for ZTF $r$ band observation   \\
\hline
	\enddata
	\tablenotetext{}{{(This table is available in its entirety a machine-readable format in the online Journal and in the China-VO PaperData Repository.)}}
\end{deluxetable*}

\section{Discussion} \label{sec:dis}
\begin{figure}[!htb]
\includegraphics[page=1,width=0.4\textwidth]{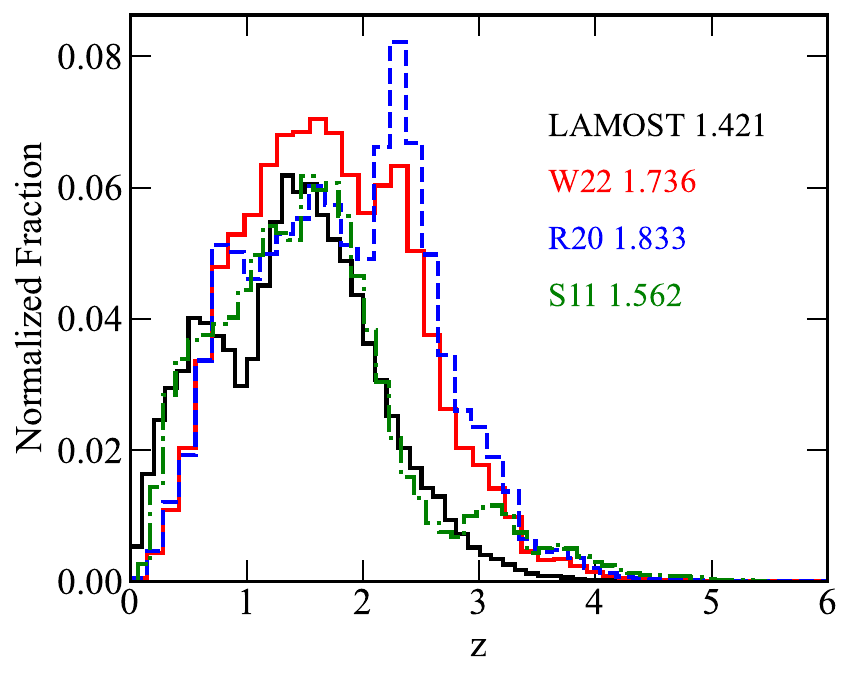}   
		\caption{The redshift distributions of LAMOST (DR1 to DR12) and  SDSS (DR16Q, DR14Q, and DR7Q) \citep[][]{2022ApJS..263...42W,2020ApJS..249...17R,2011ApJS..194...45S} quasar samples. The mean redshifts are tabulated in the right.} 
\label{fig:redshift} 
\end{figure}

\begin{figure}[!htb]
\includegraphics[page=1,width=0.4\textwidth]{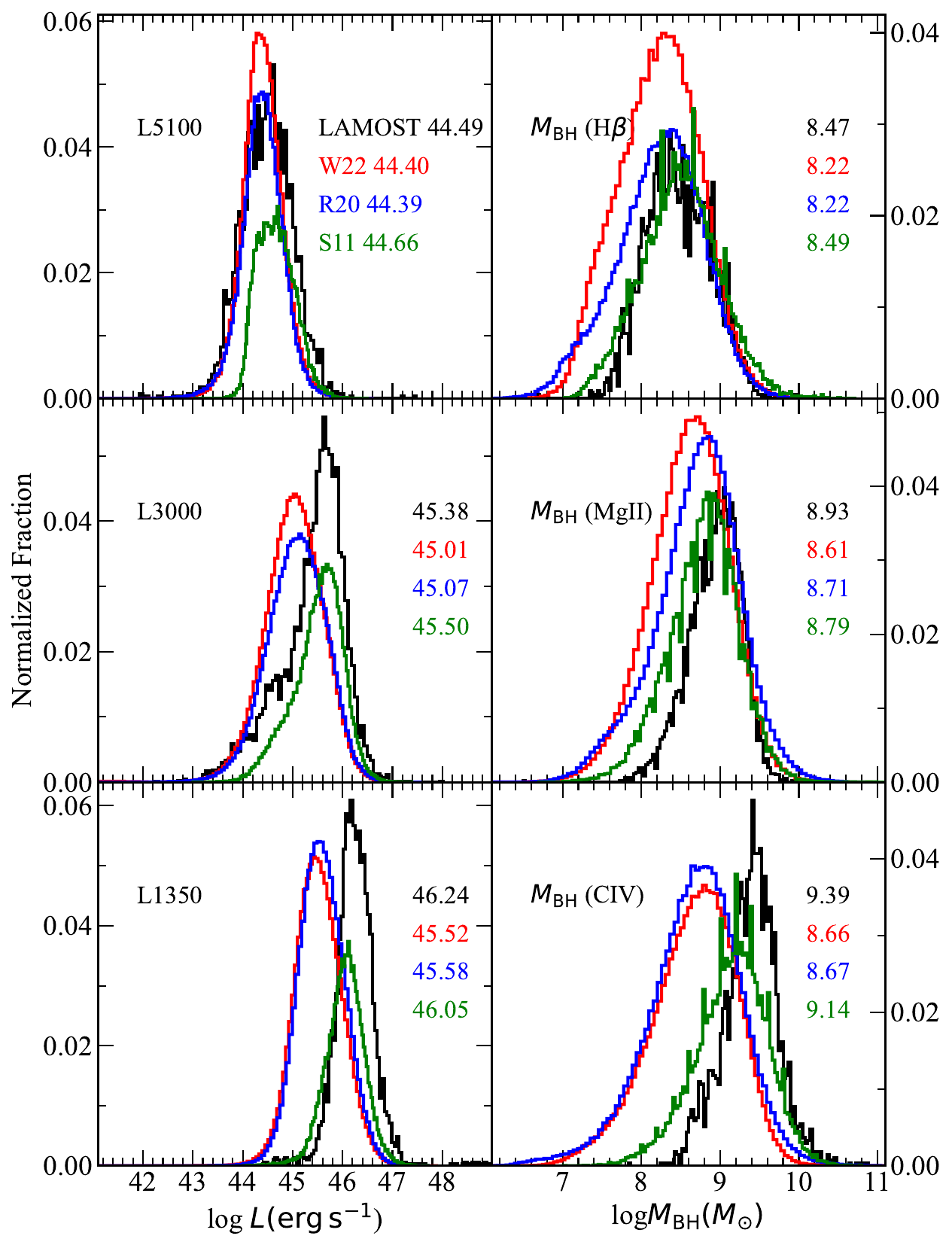}   
		\caption{The histograms of the monochromatic continuum luminosities ($L_{5100}$, $L_{3000}$, and $L_{1350}$) and the estimated $\rm M_{BH}$  (based on H$\beta$, Mg{\sc ii} and C{\sc iv}) for LAMOST and  SDSS \citet{2022ApJS..263...42W,2020ApJS..249...17R,2011ApJS..194...45S} quasar sample. The mean value of each distribution is tabulated in the right. } 
\label{fig:L_M} 
\end{figure}

We compare the basic properties of quasars identified in LAMOST DR1–DR12 to those from SDSS. \autoref{fig:redshift} presents the redshift distribution for LAMOST and different SDSS quasar Data Release samples. The mean redshift of LAMOST quasars is slightly lower than that of SDSS. \autoref{fig:L_M} presents the distributions of the monochromatic continuum luminosities ($L_{5100}$, $L_{3000}$, $L_{1350}$) and the $\rm M_{BH}$ based on H$\beta$, Mg{\sc ii} and C{\sc iv}. For the monochromatic continuum luminosities of the LAMOST sample, $L_{5100}$ is similar to \hyperlink{cite.2022ApJS..263...42W}{W22} and \hyperlink{cite.2020ApJS..249...17R}{R20}, and slightly lower than \hyperlink{cite.2011ApJS..194...45S}{S11}. $L_{3000}$ and $L_{1350}$ are similar to \hyperlink{cite.2011ApJS..194...45S}{S11} and higher than \hyperlink{cite.2022ApJS..263...42W}{W22} and \hyperlink{cite.2020ApJS..249...17R}{R20}. For the $\rm M_{BH}$ of the LAMOST quasar sample,  the distribution of $\rm M_{BH}$ is similar to \hyperlink{cite.2011ApJS..194...45S}{S11} and slightly higher than \hyperlink{cite.2022ApJS..263...42W}{W22} and \hyperlink{cite.2020ApJS..249...17R}{R20}. 

\begin{figure*}[htb]
\centering
\includegraphics[page=1,width=0.7\textwidth]{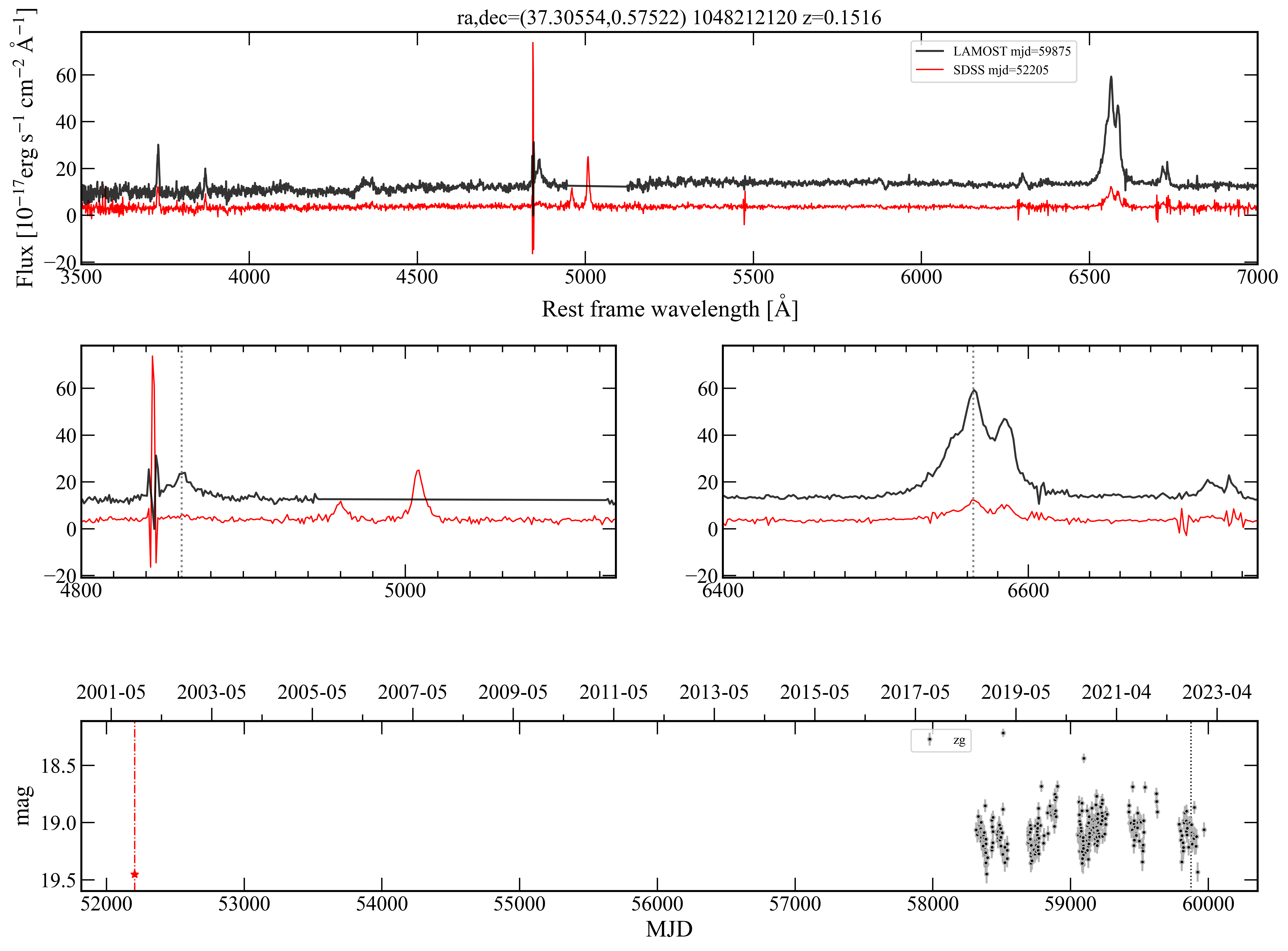}   
		\caption{An example of CLAGN candidates selected from the EVQ sample. The top panel presents the comparison for the spectra from LAMOST and SDSS. The middle panel presents zoomed regions for H$\beta$, and H$\alpha$ emission lines. The broad H$\beta$ component is evident in the LAMOST spectrum. The bottom panel presents the g band light curve from ZTF. The red and black vertical lines correspond to observational time for SDSS and LAMOST spectra.} 
\label{fig:CLAGNc1} 
\end{figure*}

We estimate the maximum change in the ZTF g-band magnitude to search for variable quasars. After the selection of photometric uncertainties less than $0.15$ mag and removal of outliers in the light curves, 807 sources in the LAMOST quasar DR10–DR12 catalog are found to be EVQs with a ZTF g band magnitude variability $|\Delta g|_{max} >1$ mag. EVQs are ideal candidates to search for changing-look AGNs (CLAGNs) with the appearance/disappearance of broad emission lines \citep{2016MNRAS.457..389M,2018ApJ...862..109Y}. Hunting for CLAGNs by combining multi-epoch spectra from LAMOST and SDSS has been reported in \citet{2018ApJ...862..109Y,2025ApJ...986..160D,2025ApJ...980...91Y}. \citet{2018ApJ...862..109Y} found 21 new CLAGNs through various ways, including the repeating spectra from archival data of LAMOST and SDSS, photometric variability, and new spectroscopic observations.  \cite{2025ApJ...986..160D} utilized data from LAMOST catalog DR10 and DR11 and SDSS DR18 to search for CLAGNs and found 51 (40 new) CLAGNs.  \citet{2025ApJ...980...91Y} reported 82 (70 new) turn-on CLAGNs that were previously classified as galaxies and experienced significant brightening in the optical and mid-infrared bands, which are confirmed by spectroscopic survey and follow-up observations.  We present an example of a CLAGN candidate found in this work in \autoref{fig:CLAGNc1}. We can enlarge the CLAGN samples by combining the multi-epoch spectra from LAMOST and DESI \citep[e.g.,][]{2024ApJS..270...26G} in future work. 

\begin{figure*}[htb]
\centering
\includegraphics[page=1,width=0.7\textwidth]{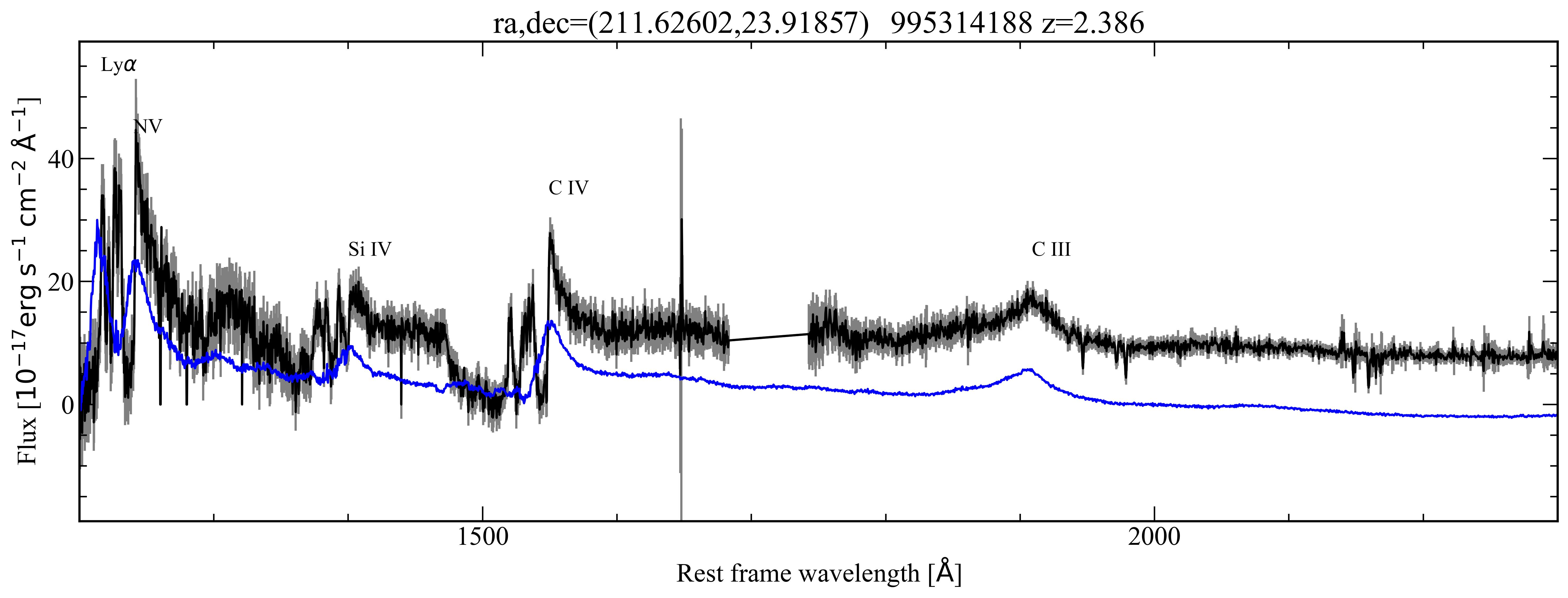}   
		\caption{An example for the $\rm Ly\alpha$ BAL quasar spectrum (black) in the rest frame. The absorption features in the $\rm Ly \alpha$,  $\rm Si$\,{\sc iv}, and C{\sc iv} are obvious. The blue line is the BAL template spectra for comparison at the same redshift. There is no correction for the Galactic extinction.} 
\label{fig:BAL} 
\end{figure*}

Except for CLAGNs, LAMOST spectra could help us search for and study other kinds of rare quasars. Quasars with broad absorption lines complicate studies of large-scale structure and cosmological studies \citep{2024MNRAS.532.3669F}. We find 110 BALs that show broad absorption features at Mg{\sc ii} and/or C{\sc iv} in this work. Thus, a total of 400 BALs have been reported in the LAMOST quasar survey (57 BALs in Paper \hyperlink{cite.2019ApJS..240....6Y}{III}, and 233 BALs in Paper \hyperlink{cite.2023ApJS..265...25J}{IV}). { The fraction of BAL quasars in the LAMOST quasar survey is systematically less than the normal fraction of optically selected quasar samples, which is mainly due to the low S/N of LAMOST spectra.} We present an example spectrum of a $\rm Ly\alpha$ BAL quasar in \autoref{fig:BAL}. 

BFFs are a special type of transient and rarely occur in broad line AGNs \citep{2019NatAs...3..242T,2023ApJ...953...32M}. \citet{2023ApJ...957...57D} found one BFF from the rising events based on the ZTF public alerts, which occurred within a galaxy nucleus and was classified by a spectroscopic approach. The strong \NIII\ and \OIIIbf\ emission lines (BF lines) produced by Bowen Fluorescence are indicators of extreme ultraviolet radiation and might be associated with the BLR and accretion process. \citet{2023ApJ...953...32M} argue that a late-time rebrightening after a flare seems to be another intriguing common property for BFFs. Combining the LAMOST spectra and ZTF light curve, we aim to search for more BFFs with both evident BF lines and multiple flare events, and further study the mechanism of such special nuclear transients. 

Follow-up observations and multi-wavelength studies of these interesting rare quasars (such as EVQ, CLAGN, BAL, and BFF, etc) from the LAMOST quasar survey should be paid more attention to understand their physical processes. Meanwhile, the ongoing LAMOST survey will continue the searches for GPQs \citep{2021ApJS..254....6F,2022ApJS..261...32F,2025ApJS..278....6H}, which would provide valuable applications in constructing the reference frame for astrometry and probing the chemistry and kinematics of the interstellar/intergalactic medium of the Milky Way by using GPQs as backlights.

\section{Summary} \label{sec:sum}
In this work, we present the LAMOST quasar survey in the tenth, eleventh, and twelfth Data Release (DR10-12). There are $11,346$ visually inspected quasars, among which  $5,960$ are common in the Million Quasars catalog, and the remaining $5,386$ are newly discovered. After the 12-year regular survey, a total of $67,521$ quasars have been identified in the LAMOST quasar survey. $40,539$ of them are independently discovered, and $29,513$ are newly discovered. 

The spectra from the LAMOST quasar survey DR10-12 are calibrated with (quasi-)simultaneous photometric magnitude data from ZTF. The H$\alpha$, H$\beta$, Mg{\sc ii}, and  C{\sc iv} emission lines are well-fitted with the absolute flux calibrated spectra, and virial black hole masses are estimated.  These measurements are compiled into the LAMOST quasar catalog and will be available online. 

The LAMOST quasar catalog will provide us with important applications in searching for rare quasars such as EVQs, CLQs, and BALs, and revealing their intrinsic physical properties. In addition, the search for GPQs using LAMOST observations will continue.

\begin{acknowledgments}
We acknowledge the anonymous referee for constructive comments and suggestions. We acknowledge the support of the Cultivation Project for LAMOST Scientific Payoff and Research Achievement. We are thankful for the support of the National Key R\&D Program of China (2025YFA1614101) and the National Science Foundation of China (12133001).  Guoshoujing Telescope (the Large Sky Area Multi-Object Fiber Spectroscopic Telescope, LAMOST) is a National Major Scientific Project built by the Chinese Academy of Sciences. Funding for the project has been provided by the National Development and Reform Commission. LAMOST is operated and managed by the National Astronomical Observatories, Chinese Academy of Sciences. Based on observations collected at the Samuel Oschin Telescope 48-inch and the 60-inch Telescope at the Palomar Observatory as part of the Zwicky Transient Facility project. The ZTF forced-photometry service was funded under the Heising-Simons Foundation grant 12540303 (PI: Graham). This research has made use of data products from the Sloan Digital Sky Survey (SDSS, \url{www.sdss.org}), the Million Quasars catalog (\url{https://quasars.org/milliquas.htm}), NASA's Astrophysics Data System (ADS, \url{https://ui.adsabs.harvard.edu/}), and NASA/IPAC EXTRAGALACTIC  DATABASE (NED, \url{https://ned.ipac.caltech.edu/}).

\end{acknowledgments}

\clearpage
\bibliography{new.ms.bib}{}
\bibliographystyle{aasjournal}

\end{document}